\documentclass[12pt,fleqn]{article}

\usepackage{bbm}
\usepackage{amsopn}
\usepackage{amsfonts}
\usepackage{amssymb}
\usepackage{amsthm}
\usepackage{amsmath}

 \DeclareMathOperator{\Mat}{Mat}

\newcommand{\tree}{\mathcal{T}}
\newcommand{\mcalL}{{\mathcal{L}}}
\newcommand{\mcalM}{{\mathcal{M}}}
\newcommand{\mcalN}{{\mathcal{N}}}
\newcommand{\mcalR}{{\mathcal{R}}}
\newcommand{\mcalP}{{\mathcal{P}}}

\renewcommand{\theequation}{\arabic{section}.\arabic{equation}}

\def\tree       {{\cal T}}

\def\curl       {\mbox{\rm curl\,}}

\newtheorem{theorem}{Theorem}[section]
\newtheorem{lemma}[theorem]{Lemma}
\newtheorem{corollary}[theorem]{Corollary}
\newtheorem{proposition}[theorem]{Proposition}
\theoremstyle{definition}
\newtheorem{definition}[theorem]{Definition}

\begin{document}

\title{\vspace{-0.75in}\Huge
Charge Superselection Sectors for Scalar QED on the Lattice}

\author{
    J. Kijowski \\
    Center for Theoretical  Physics, Polish Academy of Sciences\\
    al. Lotnik\'ow 32/46, 02-668 Warsaw, Poland\\
    \ \\
    G. Rudolph \\
    Institut f\"ur Theoretische Physik, Universit\"at Leipzig\\
    Augustusplatz 10/11, 04109 Leipzig, Germany\\
    \ \\
    C. \'Sliwa \\
    Center for Theoretical Physics, Polish Academy of Sciences\\
    al. Lotnik\'ow 32/46, 02-668 Warsaw, Poland
    }

\maketitle

\begin{abstract}
  The lattice model of scalar quantum electrodynamics (Maxwell field coupled
  to a complex scalar field) in the Hamiltonian framework is discussed.
  It is shown that the algebra of observables ${\cal O}({\Lambda})$
  of this model is a $C^*$-algebra, generated by a set of
  gauge-invariant elements satisfying the Gauss law and some additional
  relations. Next, the faithful, irreducible and non-degenerate
  representations of ${\cal O}({\Lambda})$ are found. They are labeled
  by the value of the total electric charge, leading to a
  decomposition of the physical Hilbert space into charge
  superselection sectors. In the Appendices we give a unified
  description of spinorial and scalar quantum electrodynamics
  and, as a byproduct, we present an interesting example
  of weakly commuting operators, which do not commute strongly.
\end{abstract}

\newpage

\tableofcontents

\cleardoublepage

\setcounter{equation}{0}
\section{Introduction}

The ideas of axiomatic and algebraic quantum field theory in the
sense of Wightman, Haag and Kastler \cite{HK} have played an
important role in clarifying basic nonperturbative structures of
quantum physics. In particular, there has been developed a general
scheme for superselection rules \cite{DHR}, which, however, does
not apply to theories with massless particles. Thus, an extension
of these ideas to realistic gauge theories is still a big
challenge. Some partial results in this direction already exist,
see a series of papers by Strocchi and Wightman (\cite{SW1},
\cite{S1} and \cite{S2}). In particular, in \cite{SW1} Quantum
Electrodynamics was considered. It was shown that if one insists
in locality and Lorentz covariance, one is rather naturally led to
a theory with indefinite metric. Within this scheme, the charge
superselection rule for QED was proven, but a decomposition of the
physical Hilbert space into a direct sum of subspaces carrying
definite total charge was not obtained. For a deep discussion of
charged states in QED we refer to \cite{Bu} and further references
therein. Studying simple toy models, e.g. a $Z_2$-gauge theory
with $Z_2$-matter fields \cite{Fre}, one can realize the full
programme, which one would like to implement for realistic
theories. In \cite{Fre}, the authors were able to determine the
ground state and charged states explicitly. Using methods of
Euclidean quantum field theory, it was possible to show that --
for some regions in the space of coupling constants -- the
thermodynamic limit for charged states can be controlled.

In this paper we also discuss a simplified model, we put scalar
quantum electrodynamics on a finite lattice. In the context of
lattice approximation complicated operator theoretic problems
arising in (continuum) quantum field theory become simpler,
whereas problems typical for gauge theories remain and can be,
therefore, discussed separately. We consider scalar QED in the
hamiltonian approach on a finite cubic (3-dimensional) lattice and
work in the non-compact formulation, where the gauge potential
remains Lie-algebra-valued on the lattice level. Our starting
point are the commutation relations for canonically conjugate
pairs of gauge-dependent lattice fields. Then, in a first step we
construct the algebra of observables as the algebra of gauge
invariant operators fulfilling the Gauss law and show that the
charge superselection rule holds. Algebraically, the observable
algebra is generated by electric and magnetic flux operators,
together with gauge invariant operators bilinear in the matter
fields. These gauge invariant generators fulfil a number of
algebraic identities, which by using the technical tool of a
lattice tree can be reduced essentially. It turns out that the
observable algebra is the tensor product of an associative algebra
generated by canonical commutation relations (electromagnetic
part) and an associative algebra generated by a certain Lie
algebra (matter field part). Using Woronowicz's theory of
$C^*$-algebras generated by unbounded elements (see \cite{unb}),
we can endow the observable algebra with a $C^*$-structure.

Within this setting, we are able to classify all faithful,
irreducible and non-degenerate representations of the observable
algebra and to obtain the physical Hilbert space as a direct sum
of representation spaces labeled by the total charge. We stress
that the restriction to non-degenerate representations of
$C^*$-algebras (or, equivalently, to {\em integrable}
representations of the underlying Lie algebras) is of fundamental
importance, as is shown in Appendix \ref{Example}.

We have obtained a similar result for the case of spinorial QED
earlier, see \cite{qedspin} and \cite{qedspin1}. However, from the
mathematical point of view, the problems occurring in this paper
are much more complicated due to the fact that the matter field
part of the observable algebra is generated by a non-compact Lie
algebra. Consequently, the observable algebra is
infinite-dimensional and its representations are more difficult to
control. For some earlier work on scalar lattice QED we refer to
\cite{qedscal} and for basic notions concerning lattice gauge
theories we refer to \cite{Seiler} and references therein.
Recently, we have started to investigate lattice QCD in the above
spirit, see \cite{KRQCD}.

Our paper is organized as follows: In Sections
\ref{LatticeFormulation} we discuss the standard second
quantization procedure on the lattice. We define the field
algebra, discuss the Gauss law and introduce the notion of
boundary data. In Section \ref{Observablealgebra} we construct
observables in terms of field operators and give an abstract
definition of the observable algebra in terms of generators and
relations. Finally, we endow it with a $C^*$-structure. In Section
\ref{RepresentationsSSS} we find all faithful, irreducible and
non-degenerate representations of the observable algebra  and
prove that they are labeled by the eigenvalues of the total charge
operator. This yields the superselection structure. Finally, we
discuss some perspectives of our approach. In Appendix
\ref{Example} we give an example of a non-integrable
representation carrying non-integer charge and in Appendix
\ref{Unification} we present a unified description for both
bosonic and fermionic matter.

\section{Scalar QED on the Lattice}
\label{LatticeFormulation}
\setcounter{equation}{0}

Continuum scalar quantum electrodynamics (QED) is the theory of a
comp\-lex-valued scalar field $\phi$ interacting with the
electromagnetic field $A_\mu$. The classical Lagrangian of this
model is defined as follows
\begin{equation}
 \mathcal{L} = - \frac{1}{4} F_{\mu\nu} F^{\mu\nu}
 - \frac{1}{2} \overline {D_\mu \phi} D^\mu \phi
 - V(|\phi|^2),
\end{equation}
where $D^\mu \phi = \partial_\mu \phi + i g A_\mu \phi$,
$F_{\mu\nu} = \partial_\mu A_\nu - \partial_\nu A_\mu$, $g = e /
\hbar$. Local gauge transformations are given by
\begin{equation}
\label{gauge1}
  \tilde \phi(x) = e^{-i g \lambda(x)} \phi, \quad
  \tilde A_\mu(x)  =  A_\mu(x) + \partial_\mu \lambda(x).
\end{equation}
{}For a given Cauchy hyperplane $\Sigma = \{ t=\mbox{\rm const}
\}$ in Minkowski space, the above Lagrangian gives rise to an
infinite-dimensional Hamiltonian system in
variables~$(A_k,E^k,\phi,\pi)$ with the Hamiltonian
\begin{equation}\label{Ham}
 {\cal H} =
  \frac 12 \left( E_k\, E^k + B_k\, B^k \right)
  + \frac{1}{2} \, |\pi|^2 + \frac{1}{2} \,
  |\vec{D} \phi|^2 +  V(|\phi|^2),
\end{equation}
where $B = \curl A$ is the magnetic field, $E$ is the electric
field (the momentum canonically conjugate to $A$) and $\pi$
denotes the momentum canonically conjugate to $\phi \ .$

Let us take a finite, regular, cubic lattice $\Lambda$ contained in $\Sigma$,
with lattice spacing $a$, and let us denote the set of n-dimensional lattice
elements by $\Lambda^{n}, n = 0,1,2,3$. Such elements are (in increasing
order of $n$) called sites, links, plaquettes and cubes. We approximate every
continuous configuration $(A_k,E^k,\phi,\pi)$ in the following way:
\begin{eqnarray}
\Lambda^{0} \ni x \longrightarrow \phi_x & := &
 \phi(x) \in \mathbbm{C} \, , \label{phix}  \\
\Lambda^{0} \ni x \longrightarrow \pi_x & := &
 \pi(x) \in \mathbbm{C} \, , \label{px}  \\
\Lambda^{1} \ni (x,x+\hat k) \longrightarrow A_{x,x+\hat k} & := &
\int_{(x,x+\hat k)} A_k\,dl \in \mathbbm{R} \,,  \label{A}  \\
\Lambda^{1} \ni (x,x+\hat k) \longrightarrow E_{x,x+\hat k} & := &
\int_{\sigma(x,x+\hat k)} E^k\,d\sigma_k \in \mathbbm{R} \,.\label{E}
\end{eqnarray}
Here $\sigma (x,x+\hat k)$ denotes a plaquette of the dual lattice, dual
to the link $(x,x+\hat k) \in \Lambda^{1}$.
A local gauge transformation of a lattice configuration is given by:
\begin{eqnarray}
 \tilde\phi_x &=& \exp(-ig\lambda_x) \, \phi_x \ , \label{gauge-phi} \\
 \tilde \pi_x &=& \exp(-ig\lambda_x) \, \pi_x \ , \label{gauge-p} \\
 \tilde A_{x,x+\hat k} &=& A_{x,x+\hat k} + \lambda_{x+\hat k} -
 \lambda_x \, ,
\label{gauge-A}
\end{eqnarray}
where $\Lambda^{0} \ni x \longrightarrow  \lambda_x \in
\mathbbm{R}$. The electric field $E$ is gauge invariant.
Note that we have chosen the non-compact lattice approximation, where the
potential and the field strength remain Lie-algebra-valued on the lattice
level.

We define second quantization of the lattice theory by postulating the
following canonical commutation relations for the lattice quantum
fields $(\hat A,\hat E,\hat \phi,\hat \pi)$ corresponding to the classical
lattice fields given by
(\ref{phix}) -- (\ref{E}):
\begin{eqnarray}
 \left[ \hat\phi^{*}_x, \hat \pi_y \right] & = &
 2 i \hbar \delta_{xy} \hat {\mathbbm{1}} =
 \left[ \hat\phi_x, \hat \pi^{*}_y \right]
  \ , \label{phip-lattice}   \\
 \left[ \hat A_{x,x+\hat k}, \hat E_{y,y+\hat l} \right] & =  &
 i\hbar\delta_{(x,x+\hat k),({y,y+\hat l})}
 \hat {\mathbbm{1}} \ . \label{EA-lattice}
\end{eqnarray}
The remaining commutators have to vanish. Here, $\delta_{(x,x+\hat
k),({y,y+\hat l})} = 0$ if $(x,x+\hat k)$ and $({y,y+\hat l})$ are
different links, $\delta_{(x,x+\hat k),({y,y+\hat l})} = 1$ if
they coincide and have the same orientation and $\delta_{(x,x+\hat
k),({y,y+\hat l})} = - 1$ if they coincide and have opposite
orientations.

All irreducible representations in the strong (Weyl) sense of the above
algebra are equivalent to the Schr\"odinger representation, see \cite{vN}, of
wave functions $\Psi \in {\cal H}_0 \equiv L^2(A,\phi) \, .$
We denote the field algebra of bounded operators on ${\cal H}_0 \, ,$ generated by
$(\hat A,\hat E,\hat \phi,\hat \pi)$ and fulfilling (\ref{phip-lattice}) and
(\ref{EA-lattice}), by ${\cal F}(\Lambda) \, .$
It is endowed with a natural $*$-operator, such that $\hat A$ and $\hat E$ are
self-adjoint. Obviously, ${\cal F}(\Lambda)$ contains a lot of unphysical
(gauge-dependent) elements. Moreover, the above electric field
$\hat E$ does not automatically satisfy the Gauss law. In what follows we will
present an explicit construction of the algebra ${\cal O}(\Lambda)$ of
observables (gauge invariant operators satisfying the Gauss law),
together with a complete classification of its irreducible
representations.

The group of local gauge transformations acts on ${\cal F}(\Lambda)$
by automorphisms, whose generators are given by
\begin{equation}
\label{Gx}
\Lambda^{0} \ni x \longrightarrow
{\hat {\cal G}}_x := -\frac i\hbar ( \sum_{\hat k}
 \hat E_{x,x+\hat k} - \hat q_x )
 \in \mbox{\rm End}( {\cal F}(\Lambda)) \ ,
\end{equation}
with
\begin{equation}
\label{ladunek}
 \hat q_x = \frac{e}{\hbar} \left( \ \mbox{\rm Im} (\hat\phi^{*}_x \ \hat \pi_x)
  - \hbar \ \hat{\mathbbm 1} \right)
\end{equation}
denoting the operator of electric charge at $x \, .$ Thus, the
corresponding (local) Gauss law constraint, which has to be imposed on
observables, has the following form
\begin{equation}
\label{GaussLaw}
\sum_{\hat k} \hat E_{x,x+\hat k} = \hat q_x  \ .
\end{equation}
We define the operator $\hat Q$ of total electric charge putting
\begin{equation}
\label{Q=j}
 \hat Q := \sum_{x\in\Lambda^{0}} \hat q_x  \ .
\end{equation}
Summing up the local Gauss laws over all lattice sites, we see
that nontrivial values of the total charge $\hat Q$ can only arise
from nontrivial boundary data, which we are now going to
introduce.  For this purpose we consider also external links of
$\Lambda$, connecting lattice sites belonging to the boundary
$\partial\Lambda$ with ``the rest of the world''. We denote these
external links by $(x,\infty) $ and consider their electric fluxes
$E_{x,\infty}$. Then we obtain from the Gauss law
\begin{equation}
\label{totalQ} \hat Q = \sum_{x\in\partial \Lambda^{0}} \hat
E_{x,\infty}  \ ,
\end{equation}
where we denote $\partial\Lambda^{n} :=
\partial\Lambda \cap \Lambda^{n}$. In this paper we assume that
the fluxes $E_{x,\infty}$ are constant in time.

For purposes of the construction of the complete field theory {\em
via} a limit of lattice approximations, we may treat $\Lambda$ as
a piece of a bigger lattice $\widetilde{\Lambda}$. Then the
boundary flux operators $\hat E_{x, \infty}$ belong to ${\cal
F}(\widetilde{\Lambda})$ and -- due to locality of the theory --
must commute with all elements of the field algebra ${\cal
F}(\Lambda)$. Physically, these external fluxes measure the
``violation of the local Gauss law'' on the boundary $\partial
\Lambda \, ,$
\begin{equation}
\label{ext}
 \hat E_{x,\infty} := {\hat q}_{x} - \sum_{\hat k}
 \hat E_{x,x+\hat k}  \ .
\end{equation}
This is due to the fact, that the ``world outside of $\Lambda$''
has been discarded on this level of approximation. According to
the above discussion, we assume that the above elements belong to
the center of the algebra ${\cal F}(\Lambda)$. Mathematically,
admitting non-vanishing elements of this type is equivalent to
admitting gauge dependence of quantum states under the action of
boundary gauges $\partial \Lambda^0 \ni x  \rightarrow \xi(x) \in
U(1) \, . $

As will be shown, the charge operator $\hat Q$ defines a
superselection rule, giving $\hat Q = Q \hat{\mathbbm{1}} $ on
every superselection sector. Consequently, the only consistent
choice for the external fluxes is $\hat E_{x,\infty} =
E_{x,\infty} \, \hat{\mathbbm{1}}$ on every superselection sector,
where $E_{x,\infty}$ are $c$-numbers fulfilling
\begin{equation} \label{total}
 Q = \sum_{x\in\partial \Lambda^{0}} E_{x,\infty}  \ .
\end{equation}
Therefore, we treat external fluxes as prescribed, classical
boundary conditions and show that representations characterized by
the same value $Q$, but corresponding to different external flux
distributions fulfilling (\ref{totalQ}) are equivalent. For a more
detailed discussion of this point we refer to \cite{KRQCD}.

When considering $\Lambda$ as a piece of a bigger lattice
$\widetilde{\Lambda}$, we must also prescribe the interaction of
the magnetic degrees of freedom on $\Lambda$ with the rest of the
world: in continuum theory, an additional condition for $B^\|$ or
$B^\perp$ on the boundary is necessary. In lattice theory, these
quantities live on external plaquettes of lattice cubes, adjacent
to $\partial\Lambda$. In what follows, we simply assume the
boundary condition $B^\| = 0$ over the whole boundary
$\partial\Lambda$. Due to the Maxwell equation $$\dot E^\perp =
\mbox{\rm curl}_2 B^\|\ ,$$ this condition is compatible with the
fact that $\hat E_{x,\infty}$ are time-independent.  We stress
that other boundary conditions could be considered as well.

\section{The Observable Algebra} \label{Observablealgebra}
\setcounter{equation}{0}

The observable algebra ${\cal O}(\Lambda )$ will be defined by
imposing the local Gauss law and gauge invariance. To implement
gauge invariance we have to take those elements of ${\cal
F}(\Lambda )\, ,$ which commute with all generators $\hat {{\cal
G}_x} \,.$ In Subsection \ref{generators} we give a complete list
of gauge invariant generators, built from elements of the field
algebra. These generators are not independent, they have to fulfil
a number of relations. Moreover, as an additional relation we
impose the Gauss law. We define ${\cal O}(\Lambda )$ as a
$C^*$-algebra generated by unbounded elements, fulfilling these
relations.

\subsection{Generators and defining relations}
\label{generators}

We start with giving a complete list of generators of ${\cal
O}(\Lambda )$ on a purely algebraic level. Obviously, the electric
field $$ \Lambda^{1} \ni (x,x+\hat k) \rightarrow \hat E_{x,x+\hat
k} \in \mathbbm{R} \, , $$ as well as the magnetic flux through
the rectangular \emph{plaquette} $(x;\hat k,\hat l)$ at~$x \, ,$
spanned by the vectors $\hat k$ and~$\hat l \, ,$ $$ \Lambda^{2}
\ni (x;\hat k,\hat l)\rightarrow \hat B_{x;\hat k,\hat l} \in
\mathbbm{R} \, , $$ defined by
\begin{eqnarray}\label{fluxB}
 \hat B_{x;\hat k,\hat l} =
 \hat A_{x,x+\hat k} + \hat A_{x+\hat k,x+\hat k+\hat l} +
 \hat A_{x+\hat k+\hat l,x+\hat l} + \hat A_{x+\hat l,x}
\end{eqnarray}
are gauge invariant. With every lattice path~$\gamma ,$ starting
at $x$ and ending at $y$, we associate the following set of
generators:
\begin{eqnarray}\label{l-def}
 \hat {\mcalL}_\gamma & = & \hat \phi_x^{*}
 \exp(i g \int_\gamma \hat A) \hat \phi_y, \\
 \hat {\mcalM}_\gamma & = & \hat \phi_x^{*}
 \exp(i g \int_\gamma \hat A) \hat \pi_y, \\
 \hat {\mcalN}_\gamma & = & \hat \pi_x^{*}
 \exp(i g \int_\gamma \hat A) \hat \phi_y, \\
 \hat {\mcalR}_\gamma & = & \hat \pi_x^{*}
 \exp(i g \int_\gamma \hat A) \hat \pi_y \, .\label{r-def}
\end{eqnarray}
We will use the symbol $\hat {\mcalP}_\gamma$ as a place holder
for any of $\hat {\mcalL}_\gamma$, $\hat {\mcalM}_\gamma$, $\hat
{\mcalN}_\gamma$ or $\hat {\mcalR}_\gamma$. It is clear that the
set $(\hat B, \hat E,\hat {\mcalP}_\gamma)$ generates the
observable algebra.

Gauge invariance of the theory, together with canonical
commutation relations fulfilled by elements of field algebra
generators $(\hat A,\hat E,\hat \phi,\hat \pi)$, impose many
relations between generators $(\hat B, \hat E,\hat
{\mcalP}_\gamma)$ of the observable algebra. The proposition below
contains a set of relations, which is minimal in the following
sense: Taking them, together with the Gauss law, as defining
relations of ${\cal O}(\Lambda )$, we show that its faithful
irreducible representations are unique, for a given value of the
total charge.

\begin{proposition}
\label{minset} The following relations between generators of the
observable algebra hold:
\begin{enumerate}
\item
For any lattice site $x$ we have the Gauss law:
\begin{equation}
  \label{eq:gl}
  \sum_{\hat k} \hat E_{x, x+\hat k} =
  \frac{e}{\hbar} \left( \ \mbox{\rm Im} (\hat {\cal M}_{xx})\right)
  - e \ \hat{\mathbbm 1}       \, ,
\end{equation}
where the index ``$xx$'' on the right hand side stands for the
trivial path at $x$.
\item
For any lattice site $x$ and three independent vectors $(\hat
k,\hat l,\hat n ) \, ,$ spanning a lattice cube at $x$, we have
the Bianchi identity:
\begin{equation}
 \hat B_{x;\hat k,\hat l} + \hat B_{x;\hat n,\hat k} +
 \hat B_{x;\hat l,\hat n} + \hat B_{x+\hat n;\hat l,\hat k} +
 \hat B_{x+\hat l;\hat k,\hat n} +
 \hat B_{x+\hat k;\hat n,\hat l} =  0 \ .
 \label{divB}
\end{equation}
\item
For two different lattice paths $\gamma_1$ and~$\gamma_2 \, ,$
both from~$x$ to~$y \, ,$ we have
\begin{equation}
\label{eq:twopaths}
\hat {\mcalP}_{\gamma_2} =
    \exp(i g \hat B_{\gamma_1^{-1} \gamma_2}) \hat
    {\mcalP}_{\gamma_1} \, ,
\end{equation}
where $\gamma_1^{-1} \gamma_2$ is the loop composed of
$\gamma_1^{-1}$ and~$\gamma_2$ ($\gamma_2$ is adjoint to the end
of~$\gamma_1^{-1}$), and $\hat  B_{\gamma} = \int_{\gamma} \hat A
\, .$
\item
The following commutation relations between the electromagnetic
and the matter field generators hold:
\begin{eqnarray}
  \left[ \hat{\mcalP}_\gamma, \hat B_{x;\hat k,\hat l}
  \right] &=& 0 \ ,   \label{mB} \\
  \left[ \hat{\mcalP}_\gamma, \hat E_{x,x+\hat k}
  \right] &=& e\hbar \delta_{\gamma,(x,x+\hat k) }
  \hat{\mcalP}_\gamma \ , \label{mE} \\
 \left[\hat B_{x;\hat k,\hat l}, \hat E_{y,y+\hat n} \right]
 &=& i\hbar\delta_{\partial({x;\hat k,\hat l}),(y,y+\hat n) } \ .
\label{commBE}
\end{eqnarray}
Here $\delta_{\gamma,(x,x+\hat k)} = 0$ if $(x,x+\hat k) \not\in
\gamma$ , $\delta_{\gamma,(x,x+\hat k)} = 1$ if $(x,x+\hat k) \in
\gamma$ and has the same orientation as $\gamma$ and
$\delta_{\gamma,(x,x+\hat k)} = - 1$ otherwise. In the last
formula, $\partial({x;\hat k,\hat l})$ denotes the boundary of the
oriented plaquette $(x;\hat k,\hat l)$.
\item
The following commmutation relations between the matter field
generators hold: If $\gamma$ is a path from~$x$ to~$y$, and
$\gamma'$ is a path from~$x'$ to~$y' \, ,$ then we have
\begin{eqnarray}\label{PPgamma1}
  [\hat{\mcalL}_\gamma, \hat{\mcalL}_{\gamma'}] & = & 0 \\ {}
  [\hat{\mcalL}_\gamma, \hat{\mcalM}_{\gamma'}] & = &
    2 i \hbar \delta_{x y'} \hat{\mcalL}_{\gamma'\gamma} \\ {}
  [\hat{\mcalL}_\gamma, \hat{\mcalN}_{\gamma'}] & = &
    2 i \hbar \delta_{x' y} \hat{\mcalL}_{\gamma\gamma'} \\ {}
  [\hat{\mcalL}_\gamma, \hat{\mcalR}_{\gamma'}] & = &
    2 i \hbar (\delta_{x y'} \hat{\mcalN}_{\gamma'\gamma}
    + \delta_{x' y} \hat{\mcalM}_{\gamma\gamma'}) \\ {}
  [\hat{\mcalM}_\gamma, \hat{\mcalM}_{\gamma'}] & = &
    2 i \hbar (\delta_{x y'} \hat{\mcalM}_{\gamma'\gamma}
    - \delta_{x' y} \hat{\mcalM}_{\gamma\gamma'}) \\ {}
  [\hat{\mcalM}_\gamma, \hat{\mcalN}_{\gamma'}] & = & 0 \\ {}
  [\hat{\mcalM}_\gamma, \hat{\mcalR}_{\gamma'}] & = &
    2 i \hbar \delta_{x y'} \hat{\mcalR}_{\gamma'\gamma} \\ {}
  [\hat{\mcalN}_\gamma, \hat{\mcalN}_{\gamma'}] & = &
    2 i \hbar (\delta_{x' y} \hat{\mcalN}_{\gamma\gamma'}
    - \delta_{x y'} \hat{\mcalN}_{\gamma'\gamma}) \\ {}
  [\hat{\mcalN}_\gamma, \hat{\mcalR}_{\gamma'}] & = &
    2 i \hbar \delta_{x' y} \hat{\mcalR}_{\gamma\gamma'} \\ {}
  [\hat{\mcalR}_\gamma, \hat{\mcalR}_{\gamma'}] & = & 0 \, .
  \label{PPgamma10}
\end{eqnarray}
Thus, the invariant fields $\hat{\mcalP}_\gamma$ generate a Lie
algebra.
\item
The $*$-operator acts on the matter field generators as follows:
\begin{equation}
  \label{eq:star}
  (\hat{\mcalL}_\gamma)^{*} = \hat{\mcalL}_{\gamma^{-1}}, \;
  (\hat{\mcalM}_\gamma)^{*} = \hat{\mcalN}_{\gamma^{-1}}, \;
  (\hat{\mcalN}_\gamma)^{*} = \hat{\mcalM}_{\gamma^{-1}}, \;
  (\hat{\mcalR}_\gamma)^{*} = \hat{\mcalR}_{\gamma^{-1}}.
\end{equation}
\item
For any path $\gamma$ from $x$ to $y$ and any path $\gamma^\prime$
from $y$ to $z \, ,$ the following identity holds:
\begin{equation}
\label{LLLL1}
  \hat{\mcalL}_{\gamma} \hat{\mcalL}_{\gamma^\prime} =
  \hat{\mcalL}_{yy} \hat{\mcalL}_{\gamma \gamma^\prime} \ ,
\end{equation}
\end{enumerate}
\end{proposition}

\noindent
{\bf Proof:} by a number of lengthy, but simple calculations, which we leave to the
reader.
\qed

In particular, equation (\ref{eq:gl}) follows from
equation (\ref{ladunek}), with the right-hand-side
expressed in terms of the generators
$\hat {\cal M}$,
\begin{equation}
  \label{ladunek1}
  \hat q_x = \frac{e}{\hbar}  \ \mbox{\rm Im} (\hat {\cal M}_{xx})
  - e \ \hat {\mathbbm 1}  \ .
\end{equation}
Summing over all lattice points yields the expression for the
global charge:
\begin{equation}
  \label{charge}
  \hat Q =  \frac e{\hbar} \sum_{x \in \Lambda^0}
  \ \mbox{\rm Im} (\hat {\cal M}_{xx})-  N e \ \hat{\mathbbm 1}
      \, ,
\end{equation}
with $N$ being the number of lattice points.

Now we are able to give an abstract definition of the observable algebra,
which does not refer any more to the field algebra we started with.

\begin{definition}
\label{abstrobsalg} The observable algebra ${\cal O}(\Lambda)$ is
a $C^*$-algebra generated by abstract elements $(\hat B, \hat
E,\hat {\mcalP}_\gamma) \, .$ The generators satisfy the following
axioms:
\begin{enumerate}
\item
The Gauss law (\ref{eq:gl}).
\item
The Bianchi identities (\ref{divB}).
\item
Identities (\ref{eq:twopaths}), relating generators
$\hat{\mcalP}_\gamma$ for two different paths with common end
points.
\item
The commutation relations (\ref{mB}) -- (\ref{commBE}) between
electromagnetic and matter field generators.
\item
The Lie algebra commutation relations (\ref{PPgamma1}) --
(\ref{PPgamma10}) among the matter field generators.
\item
The generators $\hat E$ and $\hat B$ are self-adjoint, $\hat
E^*=\hat E$ and $\hat B^*=\hat B \, ,$ whereas the matter field
generators fullfil (\ref{eq:star}).
\item
There is a single pair $(\gamma_0,\gamma^\prime_0)$ of non-trivial
paths with a common intermediate point $y_0 \, ,$  $\gamma_0$
connecting $x_0$ with $y_0 \ne x_0$ and $\gamma^\prime_0$
connecting $y_0$ with $z_0 \ne y_0$, such that the following
identity holds:
\begin{equation}
  \label{eq:llll}
  \hat{\mcalL}_{\gamma_0} \hat{\mcalL}_{\gamma^\prime_0} =
  \hat{\mcalL}_{y_0y_0} \hat{\mcalL}_{\gamma_0 \gamma^\prime_0}\, .
\end{equation}
\end{enumerate}
\end{definition}
The notion ``$C^*$-algebra generated by abstract elements'' is
meant in the sense of Woronowicz \cite{unb} and will be explained
in Subsection \ref{funcanalytstr}.

{}From the above axioms we obtain the following basic

\begin{lemma}
\label{3.3}
\
\begin{enumerate}
\item
All generators $\hat {\mcalP}_\gamma$ are normal:
$$ [ \hat
{\mcalP}^*_\gamma , \hat {\mcalP}_\gamma ] = 0 \, .
$$
\item
For any path $\gamma$ from $x$ to $y$ and any path $\gamma^\prime$
from $y$ to $z$, the following identities hold:
\begin{eqnarray}\label{LLLL}
  \hat{\mcalL}_{\gamma} \hat{\mcalL}_{\gamma^\prime} & = &
  \hat{\mcalL}_{yy} \hat{\mcalL}_{\gamma \gamma^\prime} \ ,
  \\ \label{RRRR}
  \hat{\mcalR}_{\gamma} \hat{\mcalR}_{\gamma^\prime} & = &
  \hat{\mcalR}_{yy} \hat{\mcalR}_{\gamma \gamma^\prime} \ ,
\end{eqnarray}
If, moreover, the end $z$ of $\gamma^\prime$ differs from $y$,
then the following identity holds:
\begin{eqnarray} \label{LMLM}
  \hat{\mcalL}_{\gamma} \hat{\mcalM}_{\gamma^\prime} & = &
  \hat{\mcalL}_{yy} \hat{\mcalM}_{\gamma \gamma^\prime} \ .
\end{eqnarray}
If, moreover, also the beginning $x$ of $\gamma$ differs from $y$,
then the following identity holds:
\begin{eqnarray}\label{NMRL}
  \hat{\mcalN}_{\gamma} \hat{\mcalM}_{\gamma^\prime} & = &
  \hat{\mcalL}_{yy} \hat{\mcalR}_{\gamma \gamma^\prime}
   = \hat{\mcalR}_{\gamma \gamma^\prime} \hat{\mcalL}_{yy}
  \ .
\end{eqnarray}
\item
The element
\begin{equation}
  \hat {\cal Z} := \tfrac 1\hbar \sum_{x \in \Lambda^0}
  \mbox{\rm Im} (\hat {\cal M}_{xx}) =
  \tfrac 1{2i\hbar}\sum_{x \in \Lambda^0}
  \left(\hat {\cal M}_{xx} -
  \hat {\cal N}_{xx} \right)
\end{equation}
commutes with all elements $\hat{\mcalP}$ and, therefore, belongs
to the center of the observable algebra.
\end{enumerate}
\end{lemma}

\noindent {\bf Proof:} Points 1. and 3. are easily proved by
direct inspection. To prove point 2. we have to consider two
cases: the generic one, when all the three points $x$, $y$ and $z$
are different and the non-generic one, when two of them coincide.
We begin with the generic case. Suppose that $x$ is different from
$x_0,y_0$ and $z_0 \, .$ Choose any path $\alpha$ from $x$ to
$x_0$. Acting with ${\rm ad} \hat {\cal M}_{\alpha}$ to the right
on both sides of (\ref{eq:llll}), (i.~e.~taking the commutator
with $\tfrac 1{2i\hbar} \hat {\cal M}_{\alpha}$) we obtain:
\begin{equation}\label{shift1}
  \tfrac 1{2i\hbar}\left[ \left( \hat{\mcalL}_{\gamma_0}
  \hat{\mcalL}_{\gamma^\prime_0} -
  \hat{\mcalL}_{y_0y_0} \hat{\mcalL}_{\gamma_0 \gamma^\prime_0}
  \right) ,   \hat{\mcalM}_{\alpha} \right] =
  \hat{\mcalL}_{\alpha\gamma_0}
  \hat{\mcalL}_{\gamma^\prime_0} -
  \hat{\mcalL}_{y_0y_0} \hat{\mcalL}_{\alpha\gamma_0 \gamma^\prime_0}
  =0 \ .
\end{equation}
Using identity (\ref{eq:twopaths}), we can replace the path
$\alpha\gamma_0$ by any other path from $x$ to $y_0$ and the path
$\gamma^\prime_0$ by any other path from $y_0$ to $z_0$. Acting
successively with appropriate operators ${\rm ad} \hat {\cal M}$
or ${\rm ad} \hat {\cal N}$, we may shift in the same way the
endpoints $y_0$ and $z_0$ to any other generic positions $y$ and
$z$. We stress that the above procedure holds for both generic and
non-generic (i.~e.~$x_0 = z_0$) initial position. The cases $x=y$
or $y=z$ are trivial. Finally, the only nontrivial non-generic
case of equation (\ref{LLLL}), namely $x=z$, may be obtained in a
similar way from the generic case, by acting with ${\rm ad} \hat
{\cal M}_{\beta}$, where $\beta$ is a path from $z$ to $x$. This
operation shifts point $z$ to $x$. This ends the proof of formula
(\ref{LLLL}).

Acting, in the generic case, with ${\rm ad} \hat {\cal R}_{zz}$ on
(\ref{LLLL}), we directly get a generic case of (\ref{LMLM}).
Acting on it once more with ${\rm ad} \hat {\cal R}_{xx}$, we get
a generic case of (\ref{NMRL}). Finally, acting successively two
times with ${\rm ad} \hat {\cal R}_{zz}$ on the latter identity,
we get a generic case of (\ref{RRRR}).

Non generic cases $x=z$ of (\ref{RRRR}), (\ref{LMLM}) and (\ref{NMRL}) are also
easily obtained by acting on corresponding generic cases with
${\rm ad} \hat {\cal M}_{\beta}$, where $\beta$ is a path from $z$
to $x$. This operation shifts point $x$ to $z$.
\qed

\vskip 0.5cm

\noindent{\bf Remarks:}
\begin{enumerate}
\item
As will be seen later, the real and the imaginary parts of
generators of the observable algebra are, in any physical
representation, unbounded self-adjoint operators. Writing bilinear
relations for unbounded operators is, in general, meaningless.
However, as will be seen in the sequel, part of the observable
algebra ${\cal O}({\Lambda})$ is generated from a certain Lie
subalgebra of the Lie algebra defined by formulae (\ref{PPgamma1})
-- (\ref{PPgamma10}). We will show that non-degenerate
representations of ${\cal O}({\Lambda})$ are given by unitary
representations of the corresponding Lie group (or, equivalently,
from {\em integrable} representations of the Lie algebra).
Commutation relations (\ref{PPgamma1}) -- (\ref{PPgamma10}),
together with {\em integrability} of the representation, imply
that in equations (\ref{eq:twopaths}), (\ref{eq:llll}),
(\ref{LLLL}), (\ref{RRRR}) and (\ref{NMRL}) we always multiply
{\em strongly commuting} observables. Therefore, the products used
in these equations will be always unambiguously defined. The same
argument applies also to the right-hand-side of equation
(\ref{LMLM}) and to its left-hand-side, provided $x \ne z$. The
only problem could come from the left-hand-side of (\ref{LMLM}),
for $x = z$, because the two elements $\hat{\mcalL}_{\gamma}$ and
$\hat{\mcalM}_{\gamma^\prime}$ do not commute.

Let us discuss this point in more detail:
Without any loss of generality, we may limit ourselves to the case
$\gamma^\prime = \gamma^{-1}$. Observe that we have
\begin{eqnarray}
\label{split}
  \hat{\mcalL}_{\gamma}\hat{\mcalM}_{\gamma^\prime}
  &=& \left( {\rm Re}\hat{\mcalL}_{\gamma} + i
  {\rm Im}\hat{\mcalL}_{\gamma}\right) \cdot
  \left( {\rm Re} \hat{\mcalM}_{\gamma^\prime} + i
  {\rm Im} \hat{\mcalM}_{\gamma^\prime}\right) \nonumber \\
  & = &
  {\rm Re}\hat{\mcalL}_{\gamma}
  {\rm Re} \hat{\mcalM}_{\gamma^\prime} -
  {\rm Im}\hat{\mcalL}_{\gamma}
  {\rm Im} \hat{\mcalM}_{\gamma^\prime} \nonumber \\
  & + & i
  \left( {\rm Re}\hat{\mcalL}_{\gamma}
  {\rm Im} \hat{\mcalM}_{\gamma^\prime} +
  {\rm Im}\hat{\mcalL}_{\gamma}
  {\rm Re} \hat{\mcalM}_{\gamma^\prime} \right) \ .
\end{eqnarray}
It follows immediately from the commutation relations that ${\rm
Re}\hat{\mcalL}_{\gamma}$ commutes with ${\rm Im}
\hat{\mcalM}_{\gamma^\prime}$. Similarly, ${\rm Re}
\hat{\mcalM}_{\gamma^\prime}$ commutes with ${\rm
Im}\hat{\mcalL}_{\gamma}$. This means, that the imaginary part of
the above expression is unambiguously defined as a product of
commuting observables. To give a meaning also to its real part we
take the unambiguously defined identity
\begin{equation}\label{LM}
  \hat{\mcalL}_{\gamma} \hat{\mcalL}_{\gamma^\prime} =
  \hat{\mcalL}_{yy}\hat{\mcalL}_{\gamma\gamma^\prime} \ ,
\end{equation}
and act on both its sides with ${\rm ad} \hat {\cal M}_{\beta}$,
where $\beta$ is a path from $z$ to $x$. As a result we get
precisely the real part of expression (\ref{split}), which is,
therefore, {\em unambiguously} defined by the adjoint action of
self-adjoint operators on self-adjoint operators.  This shows that
the above relations of the observable algebra can be
meaningfully formulated on the level of unbounded operators.

\item
Applying successively operators ad$\hat{\mcalL}$ to identity
(\ref{eq:llll}), as we did in the above proof, we may produce a
lot of new identities. Each of them could also have been used as a
defining relation instead of (\ref{eq:llll}). In the Lemma, we have
listed  only those identities, which will be used in the sequel.
\item
By point 3 of the above Lemma, the observable $\hat {\cal Z}$
defines a superselection rule. Therefore, the physical Hilbert
space is a direct sum of charge superselection sectors,
$$
  {\cal H} = \oplus_{\alpha}^{\ } {\cal H}_{\alpha} \, ,
$$ on which $\hat {\cal Z}$ acts as ${\cal Z}_{\alpha}
\hat{\mathbbm{1}} \, .$ Due to definition (\ref{charge}), the same
is true for the global charge $\hat Q$ and, moreover, $\frac{1}{e}
Q_{\alpha} = {\cal Z}_{\alpha} - N$. As will be shown later,
${\cal Z}_{\alpha}$ may assume only integer values which proves
that also $\frac{1}{e} Q_{\alpha} $ is integer.
\end{enumerate}

\subsection{Generating the observable algebra from the tree data}

A convenient way to solve relations between generators is to
choose a {\em tree}, i.~e.~to choose a unique path connecting any
pair of lattice sites. More precisely, a tree is a pair~$(x_0,
\tree)$, where $x_0$ is a distinguished lattice site (called {\em
root}) and $\tree$ is a set of lattice links such that for any
lattice site $x$ there is exactly one path from $x_0$ to $x$, with
links belonging to~$\tree$. Suppose, we have chosen a tree. Then,
for any pair $(x,y)$ of lattice sites, there is a unique {\em
along tree} path from $x$ to $y$. Denote by $\hat{\mcalP}_{x,y}$
the generator $\hat{\mcalP}_{\gamma}$ corresponding to this path.
Due to equation (\ref{eq:twopaths}), the remaining generators
$\hat{\mcalP}$ may be expressed in terms of those, provided we
know the magnetic field $\hat B$. To choose independent quantities
among the electromagnetic generators, take any {\em off tree} link
$(x,x+\hat k)$, and denote by $\hat{\cal B}_{x,x+\hat k}$ the
magnetic flux through the surface spanned by the closed path
composed of the link $(x, x+\hat k)$ and the unique \emph{on-tree}
path from $x+\hat k$ to~$x$. It is easily seen that, solving the
Bianchi identities (\ref{divB}), starting from the root and moving
outside, we can reconstruct all the magnetic fluxes $\hat
B_{x;\hat k,\hat l}$ from $\hat{\cal B}_{x,x+\hat k}$. Among
electric fields we may also choose only those which correspond to
{\em off tree} links as independent quantities. Indeed, solving
the Gauss law (\ref{eq:gl}) for $\hat E$, starting from the
boundary and moving towards the root, with local charges $\hat
q_x$ expressed in terms generators $\hat{\mcalP}_{x,y}$ and the
boundary data $\hat E_{x,\infty}$ given, we can reconstruct all
the remaining elements $\hat E_{x,x+\hat k}$. This way we get the
following

\begin{proposition}
\label{reconstruction}
For a given tree $\tree \, ,$ the set of tree data
$$
(\hat{\mcalP}_{x,y} , \hat{\cal B}_{x,x+\hat k} , \hat E_{x,x+\hat k} ) \, , \,
\,  (x, x+\hat k) \notin \tree \, ,
$$
constitute a complete set of generators of ${\cal O}(\Lambda ) \, .$
\end{proposition}

For further details of the proof, we refer the reader to the
(completely analogous) proof for the case of QED with fermions,
contained in \cite{qedspin}.

Obviously, the tree data inherit commutation relations from
Definition \ref{abstrobsalg}. In particular, from point 4 of this
definition, we read off the following commutation relations:
\begin{equation}\label{B,E}
  [ \hat {\cal B}_{x, x+\hat k}, \hat E_{y, y+\hat l} ] =
    i \hbar \delta_{(x,y)} \delta_{ (\hat k, \hat l)}
    \hat{\mathbbm{1}},
\end{equation}
where $(x, x+\hat k) \notin \tree$, $(y, y+\hat l) \notin \tree$.
Hence, the independent electromagnetic fields fulfill the
canonical commutation relations. The corresponding associative
algebra, generated by these fields, will be denoted by ${\cal
O}^{em}_{\tree}$. Moreover, for $(z, z+\hat k) \notin \tree$ we
have:
\begin{eqnarray}
  [ \hat{\mcalP}_{xy}, \hat {\cal B}_{z, z+\hat k} ] & = & 0 \ , \\ {}
  [ \hat{\mcalP}_{xy}, \hat E_{z, z+\hat k} ] & = & 0 \ ,
  \label{P,E}
\end{eqnarray}
which means that ${\cal O}^{em}_{\tree}$ commutes with the
subalgebra ${\cal O}^{mat}_{\tree} \subset {\cal O}(\Lambda ) \,
,$ spanned by the matter field generators $\hat{\mcalP}_{xy} \, .$
This fact, together with Proposition \ref{reconstruction}, implies
that the observable algebra decomposes as follows:
\begin{equation}\label{tensor}
   {\cal O}(\Lambda ) = {\cal O}^{em}_{\tree} \otimes
   {\cal O}^{mat}_{\tree} \ .
\end{equation}

We know already from point 5 of Proposition \ref{minset},
respectively from point 5 of Definition \ref{abstrobsalg}, that
the algebra ${\cal O}^{mat}_{\tree}$ is generated by a Lie
algebra. The tree data inherit, of course, a Lie algebra
structure, which we are now going to describe. For this purpose,
let us label all the lattice sites by integers $k = 1,\dots , N \,
.$ (In what follows, it does not matter, which one among the
points $x_k$ coincides with the previously chosen tree root $x_0
\, .$) To simplify notation we shall write $\hat{\mcalP}_{kl}$
instead of $\hat{\mcalP}_{x_k,x_l} \,.$ Then, from point 5 of
Definition \ref{abstrobsalg}, we have the following commutation
relations:
\begin{eqnarray}\label{Lie1}
  \tfrac 1i[\hat{\mcalL}_{kl}, \hat{\mcalL}_{mn}] & = & 0 \\ {}
  \tfrac 1i[\hat{\mcalL}_{kl}, \hat{\mcalM}_{mn}] & = &
    2  \hbar \delta_{kn} \hat{\mcalL}_{ml} \\ {}
  \tfrac 1i[\hat{\mcalL}_{kl}, \hat{\mcalN}_{mn}] & = &
    2  \hbar \delta_{ml} \hat{\mcalL}_{kn} \\ {}
  \tfrac 1i[\hat{\mcalL}_{kl}, \hat{\mcalR}_{mn}] & = &
    2  \hbar (\delta_{kn} \hat{\mcalN}_{ml}
    + \delta_{ml} \hat{\mcalM}_{kn}) \\ {}
  \tfrac 1i[\hat{\mcalM}_{kl}, \hat{\mcalM}_{mn}] & = &
    2  \hbar (\delta_{kn} \hat{\mcalM}_{ml}
    - \delta_{ml} \hat{\mcalM}_{kn}) \\ {}
  \tfrac 1i[\hat{\mcalM}_{kl}, \hat{\mcalN}_{mn}] & = & 0 \\ {}
  \tfrac 1i[\hat{\mcalM}_{kl}, \hat{\mcalR}_{mn}] & = &
    2  \hbar \delta_{kn} \hat{\mcalR}_{ml} \\ {}
  \tfrac 1i[\hat{\mcalN}_{kl}, \hat{\mcalN}_{mn}] & = &
    2  \hbar (\delta_{ml} \hat{\mcalN}_{kn}
    - \delta_{kn} \hat{\mcalN}_{ml}) \\ {}
  \tfrac 1i[\hat{\mcalN}_{kl}, \hat{\mcalR}_{mn}] & = &
    2  \hbar \delta_{ml} \hat{\mcalR}_{kn} \\ {}
  \tfrac 1i[\hat{\mcalR}_{kl}, \hat{\mcalR}_{mn}] & = & 0
  \label{Lie10}
\end{eqnarray}
Moreover, from point 6 of this definition, we have:
\begin{equation}
  \label{eq:star1}
  (\hat{\mcalL}_{kl})^{*} = \hat{\mcalL}_{lk}, \;
  (\hat{\mcalM}_{kl})^{*} = \hat{\mcalN}_{lk}, \;
  (\hat{\mcalN}_{kl})^{*} = \hat{\mcalM}_{lk}, \;
  (\hat{\mcalR}_{kl})^{*} = \hat{\mcalR}_{lk}.
\end{equation}
These relations define a complex Lie algebra, denoted by
$\widetilde{\mathfrak{g}}^{mat}_{\tree}$, with Lie bracket $\frac
1i [ \cdot , \cdot ]$ and with conjugation ``$*$'', given by
(\ref{eq:star1}). Consider now the Lie algebra
$gl(2N,\mathbbm{C})$ (with ordinary Lie bracket $[ A,B ]=AB -
BA$), and with conjugation ``$*$'' defined as follows:
\begin{equation}
\label{conjugationgl}
A^* := - \mathbbm{1}_{(N,N)} A^\dag \mathbbm{1}_{(N,N)} \ ,
\, \, \, \, A \in gl(2N,\mathbbm{C}) \, .
\end{equation}
Here ``$\dag$'' denotes Hermitian conjugation and
\[
 \mathbbm{1}_{(N,N)} :=
 \left(
\begin{smallmatrix} \mathbbm{1}_{N}&
    0\\ 0& -\mathbbm{1}_{N}\end{smallmatrix} \right) \ ,
\]
with $\mathbbm{1}_{N}$ being the unit $(N \times N)$-matrix.

\begin{lemma}
The mapping
$$
F : \, \, \widetilde{\mathfrak{g}}^{mat}_{\tree} \, \,
\longrightarrow \, \, gl(2N,\mathbbm{C}) \, ,
$$
defined by
\begin{eqnarray}
 F(\hat{\mcalL}_{kl}) & := & \tfrac 1i\left(
  \begin{smallmatrix} -E_{kl}&
  iE_{kl}\\ i E_{kl}& E_{kl}\end{smallmatrix} \right)
  \ \ \ \ \ \ \,
 F(\hat{\mcalM}_{kl}) := \tfrac \hbar i
  \left( \begin{smallmatrix} -i E_{kl}&
  E_{kl}\\ -E_{kl}& -i E_{kl}\end{smallmatrix} \right)
  \label{identyfikacja1} \\
 F(\hat{\mcalN}_{kl}) & := & \tfrac \hbar i\left(
  \begin{smallmatrix} i E_{kl}&
  E_{kl}\\ -E_{kl}& i E_{kl}\end{smallmatrix} \right)
  \ \ \ \ \ \ \
 F(\hat{\mcalR}_{kl}) := \tfrac {\hbar^2} i
  \left( \begin{smallmatrix} -E_{kl}&
    -i E_{kl}\\ -i E_{kl}& E_{kl}\end{smallmatrix} \right)
  \ ,\label{identyfikacja2}
\end{eqnarray}
with $(E_{kl})$ being the canonical basis of $gl(N,\mathbbm{C})$,
is an isomorphism of complex Lie algebras with conjugation,
\begin{eqnarray}
F(\tfrac{1}{i} [\cal X, \cal Y]) & = & [F(\cal X) ,F(\cal Y)]
\, , \nonumber\\
F({\cal X}^*) & = & F(\cal X)^* \, ,\nonumber
\end{eqnarray}
for ${\cal X}, {\cal Y} \in \widetilde{\mathfrak{g}}^{mat}_{\tree}
\, .$
\end{lemma}

\noindent {\bf Proof:} by a lengthy, but simple calculation, which
we leave to the reader. \qed

\vspace{0.2cm}
\noindent
In what follows, we shall often omit writing $F$ and
identify $F(\hat{\cal P})$ with $\hat{\cal P}$.

We denote the real form of
$\widetilde{\mathfrak{g}}^{mat}_{\tree}$, corresponding to ``$*$''
by ${\mathfrak{g}}^{mat}_{\tree}$. The elements of this real Lie
algebra are  physical observables, spanned by the self-adjoint
elements $({\rm Re} \hat{\mcalP}_{kl},{\rm Im}\hat{\mcalP}_{kl}
)$. Under the above isomorphism, ${\mathfrak{g}}^{mat}_{\tree}$ is
identified with the real form
\[
u(N,N) := \{ A \in gl(2N, \mathbbm{C}) : A^* = A \}
\]
of $gl(2N,\mathbbm{C}) \, ,$ corresponding to the conjugation
defined by (\ref{conjugationgl}). Exponentiating $u(N,N)$ we
obtain the corresponding connected Lie group $U(N,N) \, ,$
consisting of those linear transformations of $\mathbbm{C}^{2N}$
which preserve the hermitian form defined by
$\mathbbm{1}_{(N,N)}$:
\[
U(N,N) := \{ U \in \Mat_{2N \times 2N} (\mathbbm{C}) : U^\dag
\mathbbm{1}_{(N,N)} U = \mathbbm{1}_{(N,N)} \} \ .
\]
As we will see, non-degenerate representations of ${\cal
O}^{mat}_{\tree}$ will be given by {\em intregrable}
representations of ${\mathfrak{g}}^{mat}_{\tree} \cong u(N,N)$,
this means unitary representations of the group $U(N,N)$.

Observe that due to identification (\ref{identyfikacja1}) we have
\begin{equation}
  \hat {\cal Z} = \tfrac{1}{\hbar} \sum_{k} \mbox{\rm Im} (\hat
  {\cal M}_{kk}) = i \mathbbm{1}_{(2N)}
      \, .
\end{equation}
This element generates the center $u(1)$ of $u(N,N)$. The
restriction of any {\em non-degenerate} representation of the
observable algebra to the center can be integrated to a
representation of the subgroup $ \{ \exp (i \tau
\mathbbm{1}_{(2N)}) \, , \tau \in  \mathbbm{R}^1 \} \cong U(1) \,
.$ This implies that the spectrum of $\hat {\cal Z}$ must be
integer. We conclude that the spectrum of the  charge operator
$\frac {\hat Q}{e}$, defined by formula (\ref{charge}) must be
integer, too.

Observe that equations (\ref{LLLL}) -- (\ref{NMRL}) (implied by
axiom (\ref{eq:llll})) may be rewritten in the following form: For
any triple $x_i$, $x_j$, $x_k$ of lattice points we have:
\begin{eqnarray}\label{LLLL-n}
  \hat{\mcalL}_{ij} \hat{\mcalL}_{jk} & = &
  \hat{\mcalL}_{jj} \hat{\mcalL}_{ik} \ ,
  \\ \label{RRRR-n}
  \hat{\mcalR}_{ij} \hat{\mcalR}_{jk} & = &
  \hat{\mcalR}_{jj} \hat{\mcalR}_{ik} \ .
\end{eqnarray}
If, moreover, $j \ne k$, then we have:
\begin{eqnarray}\label{LMLM-n}
  \hat{\mcalL}_{ij} \hat{\mcalM}_{jk} & = &
  \hat{\mcalL}_{jj} \hat{\mcalM}_{ik} \ .
  \end{eqnarray}
If, moreover, $i \ne j$, then we have:
\begin{eqnarray}\label{NMRL-n}
  \hat{\mcalN}_{ij} \hat{\mcalM}_{jk} & = &
  \hat{\mcalL}_{jj} \hat{\mcalR}_{ik}
  = \hat{\mcalR}_{ik} \hat{\mcalL}_{jj}
  \ .
\end{eqnarray}

\subsection{Functional analytic structure}
\label{funcanalytstr}

Now we are able to endow the observable algebra with a functional
analytic structure. Recall that ${\cal O}(\Lambda )$ is the tensor
product of ${\cal O}^{em}_{\tree}$ and ${\cal O}^{mat}_{\tree} \,
,$ see (\ref{tensor}). It turns out that both components can be
endowed with the structure of a $C^*$-algebra generated by
unbounded elements in the sense of Woronowicz, see \cite{unb}.

Since ${\cal O}^{mat}_{\tree}$ is generated by the Lie algebra
$u(N,N)$, we take the $C^*$-algebra $C^*(U(N,N))$ and factorize it
with respect to the ideal generated by relations (\ref{LLLL-n}) --
(\ref{NMRL-n}). (In fact, it is sufficient to impose only one of
these relations: the proof of Lemma \ref{3.3} shows that one of
these identities implies the remaining ones.)

The $C^*$-algebra $C^*(G)$ of any locally compact group may be
obtained as the $C^*$-comple\-tion of the group algebra of $G$. By
definition, the latter is the space $L^1(G)$ of integrable
functions on the group, with convolution providing the product
structure. The completion is taken with respect to the following
norm:
\begin{equation}\label{norm}
  \| f \| := sup_\pi \| \pi (f)\| \ ,
\end{equation}
where $f \in L^1(G)$, the supremum is taken over all
representations $\pi$ of the group and $\pi (f)$ denotes the
operator obtained by smearing the representation over the group
with the function $f$,  see \cite{Dixmier}, \cite{Ki}. We note
that $C^*(G)$ is one of the classes of examples considered by
Woronowicz. It is generated by any basis of the Lie algebra of
$G$, see \cite{Kazio} and \cite{unb}. We stress that these
generators are not elements of $C^*(G)$, they are only {\em
affiliated} in the $C^*$-sense.

Thus, take the algebra $C^*(U(N,N))$. To impose relations
(\ref{LLLL-n}) -- (\ref{NMRL-n}) on generators (more precisely,
their real and imaginary parts) we multiply them from both sides
by elements of $C^*(U(N,N))$ belonging to the common dense domain
of the generators (e.~g.~the so called {\em smooth} elements,
corresponding to functions of the class $C^\infty_0(U(N,N))
\subset L^1(U(N,N))$). This way we generate a double-sided ideal
${\cal J} \subset C^*(U(N,N))$. We define the matter algebra
${\cal O}^{mat}_{\tree}$ as the quotient:
\begin{equation}\label{quo}
{\cal O}^{mat}_{\tree} \cong C^*(U(N,N))/{\cal J} \, .
\end{equation}
It is worthwhile to notice that the same ideal may be obtained as
the space of elements whose norm vanishes, if we replace
(\ref{norm}) by the supremum over only those representations,
which fulfill additional identities (\ref{LLLL-n}) --
(\ref{NMRL-n}). Thus, ${\cal O}^{mat}_{\tree}$ could be defined
also as the completion of $L^1(U(N,N))$ with respect to the latter
norm (cf. \cite{Kazio}, \cite{unb}).

Next, we endow the electromagnetic component ${\cal
O}^{em}_{\tree}$ with a $C^*$-structure. Again, the theory of
Woronowicz can be applied. By the von Neumann theorem, all
irreducible representations of the (electromagnetic) canonical
commutation relations (\ref{B,E}) are isomorphic to the
Sch\"odinger representation. We take as ${\cal O}^{em}_{\tree}$
the $C^*$-algebra $CB(H)$ of compact operators acting on the
Hilbert space $H$ of this representation. Here, no additional
identities have to be imposed. Again, the generators $ \hat {\cal
B}_{x, x+\hat k}$ and $ \hat E_{y, y+\hat l}$ are affiliated with
the algebra.

Finally, the $C^*$-algebra ${\cal O}(\Lambda )$ is, by definition,
the {\em minimal} tensor product of ${\cal O}^{mat}_{\tree}$ with
${\cal O}^{em}_{\tree}$. Then, for elements $A$, affiliated with
${\cal O}^{mat}_{\tree}$, and $B$, affiliated with ${\cal
O}^{em}_{\tree}$, $A \otimes B$ is affiliated with ${\cal
O}(\Lambda )$.

\noindent{\bf Remarks:}
\begin{enumerate}
  \item We stress that ${\cal O}(\Lambda )$ is a $C^*$-algebra
  without identity. On the abstract level, the element
  $\hat{\mathbbm 1}$ appearing in formulae (\ref{ladunek1}) and
  (\ref{charge}) has to be understood as the identity in the
  multiplier algebra $M\left({\cal O}(\Lambda )\right)$, which is
  a subalgebra of the set of affiliated elements.
  \item  As noticed by Woronowicz
\cite{universal}, there exists an abstract definition of $C^*(G)$,
which applies to any topological group $G$, not necessarily being
a Lie group. More precisely, a $C^*$-algebra $A$ and a
homomorphism $\alpha : G \rightarrow M(A)$ of $G$ into unitary
elements of the multiplier algebra $M(A)$ is called a
$C^*$-algebra of $G$ and denoted $C^*(G)$ if the pair $(A, \alpha
)$ is universal in the following sense:  for any other such pair
$(B, \beta )$, there exists a morphism $\varphi \in Mor (A,B)$,
such that $\beta = \varphi \circ \alpha$ (for the definition of
$Mor (A,B)$ see \cite{unb}). An effective construction in case of
a locally compact group consists in taking the above mentioned
$C^*$-comple\-tion of the group algebra of $G$.
\end{enumerate}


\section{Representations of the Observable Algebra and the Charge
Superselection Structure}
\label{RepresentationsSSS}
\setcounter{equation}{0}

We are going to construct all faithful, irreducible and
non-degenerate representations of the observable algebra ${\cal
O}(\Lambda) \, .$ Concerning the electromagnetic part, the von
Neumann theorem guarantees uniqueness of representations. As for
the matter part, there is a one-to-one correspondence (c.f.
\cite{Dixmier}, \cite{Ki}) between non-degenerate representations
of the algebra $C^*(U(N,N))$ with unitary representations of the
Lie group $U(N,N)$. Thus, by Definition \ref{quo}, the
non-degenerate representations of ${\cal O}^{mat}_{\tree}$ are
given by those faithful, irreducible and integrable
representations of the Lie algebra ${\mathfrak{g}}^{mat}_{\tree}$,
which respect the additional relations (\ref{LLLL-n}) --
(\ref{NMRL-n}). To find them, we shall further reduce
${\mathfrak{g}}^{mat}_{\tree}$ to a certain Lie subalgebra
${\mathfrak{h}}^{mat}_{\tree} \subset {\mathfrak{g}}^{mat}_{\tree}
\, ,$ such that it also generates ${\cal O}^{mat}_{\tree}$ and
that the enveloping algebra of ${\mathfrak{h}}^{mat}_{\tree}$ does
not inherit any identities from the relations defining the ideal
${\cal J}$. We will show that, given a representation of
${\mathfrak{h}}^{mat}_{\tree}$ of the above type, relations
(\ref{LLLL-n}) -- (\ref{NMRL-n}) enable us to construct a unique
representation of ${\cal O}^{mat}_{\tree} \, ,$ for a given value
of total charge. Moreover, every representations of ${\cal
O}^{mat}_{\tree}$ is obtained this way.

To define the Lie subalgebra ${\mathfrak{h}}^{mat}_{\tree} \, ,$
we fix one lattice point, say $x_N$, and take only those $\hat
{\mcalL}$'s and $\hat {\mcalM}$'s, which start at this point. More
precisely, we denote:
\begin{eqnarray}
Q_k & := & \hat {\mcalL}_{Nk} \ , \\ P_k & := & \hat {\mcalM}_{Nk}
\ ,\\ R   & := & \hat {\mcalL}_{NN} \ , \\ K & := & \mbox{\rm Re}
\ (\hat {\mcalM}_{NN})
 \label{ReM}\ ,
\end{eqnarray}
for $k = 1, \dots , N-1$. Observe that $P$ and $Q$ are normal,
$[P,P^*] = [Q,Q^*] = 0 \ , $ whereas $R$ and $K$ are hermitean,
$R^* = R \ , \ K^* = K \ .$ Denoting $Q_k = q^1_k + i q^2_k$ and
$P_k = p^1_k + i p^2_k$, with $q^a_k$ and $p^a_k$ being
self-adjoint, we read off from relations (\ref{Lie1}) --
(\ref{Lie10}) the following commutation relations:
\begin{eqnarray}
  \tfrac 1i \left[ q^a_k , \tfrac 1\hbar p^b_l \right]
  & = &  \ R \  \delta_{kl} \delta^{ab}\ , \label{qp-comm}
  \\
  \tfrac 1i \left[ R , \tfrac 1\hbar K \right] & = &
  R \ , \label{rk-comm}
  \\
  \tfrac 1i \left[ q^a_k , \tfrac 1\hbar K\right] & = &
  q^a_k \ , \label{Qk-comm}
  \\
  \tfrac 1i \left[ p^a_k , \tfrac 1\hbar K\right] & = &
  p^a_k \ , \label{Pk-comm}
\end{eqnarray}
whereas the remaining commutators vanish.

Observe that formulae (\ref{qp-comm}) are the commutation
relations of the real Heisenberg Lie algebra ${\cal H}_{2(N-1)} \,
,$ generated by $2(N-1)$ canonically conjugate pairs
$(q^a_j,\tfrac 1\hbar p^a_j)$ and with center generated by $R$.
The $1$-dimensional Lie algebra generated by $\frac{1}{\hbar}K$
acts by (\ref{rk-comm}) -- (\ref{Pk-comm}) on ${\cal H}_{2(N-1)}
\, ,$ endowing  ${\mathfrak{h}}^{mat}_{\tree}$ with the structure
of a semidirect sum,
\begin{equation}
{\mathfrak{h}}^{mat}_{\tree} = {\mathbbm R}^1 \oplus_s {\cal
H}_{2(N - 1)} \ .
\end{equation}
Observe that the enveloping algebra spanned by
${\mathfrak{h}}^{mat}_{\tree}$ does not inherit any identities
from the defining relations (\ref{LLLL-n}) -- (\ref{NMRL-n}),
indeed.

\subsection{Representations of the Lie algebra
${\mathfrak{h}}^{mat}_{\tree}$}

\begin{theorem}\label{pierwsze}
There are exactly two faithful, irreducible and integrable
representations of the Lie algebra ${\mathfrak{h}}^{mat}_{\tree}$.
They are both defined on the Hilbert space ${\cal H}:= L^2
({\mathbbm C}^{N - 1} \times {\mathbbm R}^1 )$, and are given by
the following formulae:
\begin{eqnarray}\label{repr1}
  (Q_k \Psi)(z_1,\dots ,z_{N-1},\lambda) &=& \pm e^\lambda z_k
  \Psi(z_1,\dots ,z_{N-1},\lambda) \ , \\
  \label{repr2}
  (P_k \Psi)(z_1,\dots ,z_{N-1},\lambda) &=& e^\lambda
  \frac {2\hbar}{i} \frac {\partial}
  {\partial {\bar z}_k} \Psi (z_1,\dots ,z_{N-1},\lambda)\ , \\
  \label{repr3}
  (R \Psi) (z_1,\dots ,z_{N-1},\lambda) &=&
  \pm e^{2\lambda} \Psi (z_1,\dots ,z_{N-1},\lambda) \ , \\
  \label{repr4}
  (K \Psi)(z_1,\dots ,z_{N-1},\lambda) &=& \frac {\hbar}{i}
  \frac {\partial}
  {\partial \lambda} \Psi (z_1,\dots ,z_{N-1},\lambda) \ ,
\end{eqnarray}
where $\Psi \in {\cal H}$ and $(z_1,\dots ,z_{N-1},\lambda) \in
{\mathbbm C}^{N - 1} \times {\mathbbm R}^1  \, .$
\end{theorem}

\noindent {\bf Proof:} We use the following matrix representation
of ${\mathfrak{h}}^{mat}_{\tree}$:
\begin{equation}\label{X}
  X(k,r,{\bf p},{\bf q}) := \left(
\begin{array}{ccc}
  k & {\bf p} & r\\
  0 & 0 & {\bf q} \\
  0 & 0 & -k
\end{array}
\right) \, ,
\end{equation}
where ${\bf q}^T,{\bf p} \in {\mathbbm R}^{2(N-1)} \, ,$ $k,r \in
{\mathbbm R}^1 \, ,$ and the generators are identified as follows:
\begin{eqnarray}
 \tfrac{1}{\hbar} K & = & X(1,0,0,0) \, ,  \nonumber\\
  R & = & X(0,1,0,0) \, ,\nonumber\\
 \tfrac{1}{\hbar} p^1_j & = & X(0,0,\delta^i_j ,0) \, ,
 \nonumber\\
 \tfrac{1}{\hbar} p^2_j & = & X(0,0,\delta^i_{2(N-1)+j} ,0)
 \, ,\nonumber\\
 q^1_k & = & X(0,0,0,\delta^i_j) \, ,
 \nonumber\\
 q^2_k & = & X(0,0,0,\delta^i_{2(N-1)+j}) \, .\nonumber
\end{eqnarray}
Exponentiating representation (\ref{X}), we obtain the following
matrix representation of the connected, simply connected Lie group
$H^{mat}_{\tree}$, corresponding to
${\mathfrak{h}}^{mat}_{\tree}$,
\begin{equation}\label{g}
   g(u,c,{\bf a},{\bf b}) = \left(
\begin{array}{ccc}
  u & {\bf a} & c\\
  0 & {\bf 1} & {\bf b} \\
  0 & 0 & u^{-1}
\end{array}
\right) \, ,
\end{equation}
where ${\bf a}^T,{\bf b} \in {\mathbbm R}^{2(N-1)} \, ,$ $c \in
{\mathbbm R}^1$ and $u \in {\mathbbm R}_{+} \, .$ For $u = 1 \, ,$
this formula yields the standard matrix representation of the
$(4(N - 1) + 1)$-dimensional real Heisenberg group $H_{2(N-1)} \,
.$ On the other hand, for ${\bf a} = 0 = {\bf b}$ and $c=0 \, ,$
it yields the real (multiplicative) group in $1$ dimension. Thus,
$H^{mat}_{\tree}$ coincides with the following semidirect product:
\begin{equation}
H^{mat}_{\tree} =  {\mathbbm R}_{+} \times_s H_{2(N-1)} \ .
\end{equation}
Since $H_{2(N-1)}$ is a closed normal subgroup of
$H^{mat}_{\tree}$, we can apply the method of induced
representations, see \cite{Ki}. The faithful, unitary, irreducible
representations of $H_{2(N-1)}$ are labeled by $t \in {\mathbbm
R}_{*} = {\mathbbm R}^1 \setminus 0 \, ,$
\begin{equation}
\label{irepHeis}
(U_{t}(g(1,c,{\bf a},{\bf b}))f)({\bf x}) =
e^{i t({\bf b}{\bf x} + c)} \,
f({\bf x} + {\bf a}) \, ,
\end{equation}
with $f \in L^2({\mathbbm R}^{2(N-1)})$, see \cite{Ki}. Any two
representations labeled by $t$ and $t^\prime \, ,$ with $t \neq
t^\prime \, ,$ are unitarily inequivalent. We have to find the
orbits of the action of $H^{mat}_{\tree}$ on the space $\hat
H_{2(N-1)}$ of equivalence classes of unitary irreducible
representations of $H_{2(N-1)} \, .$ This action reduces to the
action of ${\mathbbm R}_{+} \, ,$ $$ \hat H_{2(N-1)} \times
{\mathbbm R}_{+} \ni (U_t,g(u,0,0,0)) \rightarrow U_t \circ {\rm
Ad} g(u,0,0,0) \in \hat H_{2(N-1)} \, . $$ Using formula
(\ref{irepHeis}) it can be shown that -- up to unitary equivalence
-- we have $$ U_t \circ {\rm Ad} g(u,0,0,0) = U_{u^2 t} \, . $$
Thus, there are two orbits, ${\cal O}_{+} = \left\{t \in {\mathbbm
R}^1 : t > 0 \right\}$ and ${\cal O}_{-} = \left\{t \in {\mathbbm
R}^1 : t < 0 \right\}$. For both orbits, the stabilizer of each
point on the orbit is conjugate to $H_{2(N-1)} \, ,$ and
${\mathbbm R}_{+}$ acts transitively on each orbit. Consequently,
there are two equivalence classes of faithful, unitary,
irreducible representations $U_{\pm}$ of $H^{mat}_{\tree}$,
induced from the representations $U_{t}\, ,$ defined on the spaces
$$ {\cal H}_{\pm} = L^2\left({\cal O}_{\pm}, L^2({\mathbbm
R}^{2(N-1)})\right) $$ of functions on ${\cal O}_{\pm}$ with
values in the representation space $L^2({\mathbbm R}^{2(N-1)})$ of
$H_{2(N-1)} \, .$ Identifying both orbits with ${\mathbbm R}^1 \,
,$ by putting $t = \pm e^{2\lambda} \, ,$ we have $ {\cal H}_{\pm}
\cong L^2({\mathbbm R}^{2(N-1)+1}) \, , $ and a simple calculation
yields the following induced representations:
\begin{equation}\label{irepg}
    (U_{\pm}(g(u,c,{\bf a},{\bf b}))f)({\bf x},\lambda) = e^{\pm i
    e^{2\lambda} \, u (e^{-\lambda}{\bf b}{\bf x} + c)} \,
    f({\bf x} + e^{\lambda}{\bf a},\lambda + \ln u ) \, .
\end{equation}
Differentiating these representations along the one parameter
subgroups generated by the above defined basis elements of  ${\cal
H}^{mat}_{\tree}$ and identifying ${\mathbbm R}^{2(N-1)}$ with
${\mathbbm C}^{(N-1)} \, ,$ we obtain exactly formulae
(\ref{repr2}) -- (\ref{repr4}). \qed

Observe that the transformation
\begin{eqnarray}\label{trans1}
  I \left(\hat{\mcalL}_{xy} \right) &=& - \hat{\mcalR}_{xy} \ , \\
  \label{trans2}
  I \left(\hat{\mcalR}_{xy} \right) &=& - \hat{\mcalL}_{xy} \ , \\
  \label{trans3}
  I \left(\hat{\mcalM}_{xy} \right) &=&  \hat{\mcalN}_{xy} \ , \\
  \label{trans4}
  I \left(\hat{\mcalN}_{xy} \right) &=&  \hat{\mcalM}_{xy} \ ,
\end{eqnarray}
preserves the defining relations (\ref{PPgamma1}) -- (\ref{RRRR})
and, therefore, generates an automorphism of ${\cal
O}^{mat}_{\tree}$. This automorphism intertwines the two
representations given by Theorem  \ref{pierwsze}. Thus, it is
sufficient to consider the positive representation only, because
the other one is equivalent to one obtained from the positive
representation by relabeling of the elements of ${\cal
O}^{mat}_{\tree}$. Such a relabeling changes, however, the
physical meaning of some observables. This is, in particular, true
for the electric charge. Indeed, the sign of $\mbox{\rm Im} (\hat
{\cal M}_{xx})$ changes under the transformation $I$.
Consequently, the definition (\ref{ladunek1}) of the electric
charge, would have to be supplemented by the sign of the
representation. Therefore, we choose the positive representation
once for ever. This choice implies the positivity of all the
elements $\hat{\mcalL}_{xx}$ and $\hat{\mcalR}_{xx}$.

It is interesting to observe that we could postulate the
positivity of only one of them, say $\hat{\mcalL}_{x_0x_0}$, as an
additional axiom for the observable algebra. Indeed, identity
(\ref{eq:llll}) implies
\[
  \hat{\mcalL}_{x_0x_0} \cdot \hat{\mcalL}_{yy} =
  \hat{\mcalL}_{x_0y} \cdot
  \hat{\mcalL}_{yx_0} =
  \hat{\mcalL}_{x_0y} \cdot \left( \hat{\mcalL}_{x_0y}
  \right)^* > 0 \ .
\]
Similarily, we get for $x \ne y$
\[
  \hat{\mcalR}_{xx} \cdot \hat{\mcalL}_{yy} = \hat{\mcalN}_{xy} \cdot
  \hat{\mcalM}_{yx} = \hat{\mcalN}_{xy} \cdot \left( \hat{\mcalN}_{xy}
  \right)^* > 0 \ .
\]
Hence, choosing the positive sign of $\hat{\mcalL}_{x_0x_0}$ we
obtain positivity for all of them.

\subsection{Local charge operators}

\begin{lemma}\label{charge1}
In the positive representation we have
\begin{equation}\label{q_k}
  (\hat q_{x_k} \Psi)(z_1,\dots ,z_{N-1},\lambda) =
  \frac ei \frac {\partial}{\partial \  \mbox{\rm arg} z_k}
  \Psi (z_1,\dots ,z_{N-1},\lambda) \ ,
\end{equation}
for $k = 1,\dots ,N-1$.
\end{lemma}

\noindent {\bf Proof:} Due to identity
\[
Q^*_k  P_k = \hat {\mcalL}_{kN} \hat {\mcalM}_{Nk} = \hat
{\mcalM}_{kk} \hat {\mcalL}_{NN} = R \hat {\mcalM}_{kk} \ ,
\]
and to formulae (\ref{repr1}) -- (\ref{repr4}), we have:
\begin{equation}\label{QP}
  R \hat {\mcalM}_{kk} = R {\bar z_k}
  \frac {2\hbar}{i} \frac {\partial}
  {\partial {\bar z_k}} \ .
\end{equation}
The operator R is strictly positive and, hence, we may skip it on
both sides of the equation. This way we obtain:
\begin{equation}\label{QP1}
  \hat {\mcalM}_{kk} = {\bar z_k}
  \frac {2\hbar}{i} \frac {\partial}
  {\partial {\bar z_k}} \ .
\end{equation}
The imaginary part of this operator gives us (\ref{q_k}).
\qed

\vspace{0.2cm} \noindent {\bf Remark :} Deriving (\ref{QP1}) from
(\ref{QP}), we have, in fact, divided both sides of the equation
by $ R = \hat {\mcalL}_{NN}$. Such an operation is, of course,
meaningless on the level of the abstract algebra ${\cal
O}^{mat}_{\tree}$. On the level of representations, however, we
have proved that $R$ cannot be a divisor of zero. Hence, this
operation is fully justified.

\begin{corollary}
In every representation of ${\cal O}^{mat}_{\tree}$, the local
charge operators $\hat q_{x_k}$ take integer eigenvalues (in units
of the elementary charge) only. Thus, the spectrum of each $\hat
q_{x_k}$ is given by
\[
\mbox{\rm Sp} (\hat q_{x_k}) = e {\mathbbm N} \ .
\]
\end{corollary}

\subsection{Constructing representations of ${\cal O}^{mat}_{\tree}$
from representations of ${\mathfrak{h}}^{mat}_{\tree}$}

We will show that each irreducible, faithful and non-degenerate
representation of ${\cal O}^{mat}_{\tree} \, ,$ assigned to a
given tree, is uniquely generated by a corresponding
representation of its Lie subalgebra
${\mathfrak{h}}^{mat}_{\tree}$, provided the total electric charge
$Q$ carried by the matter field is fixed.

Suppose that a representation of ${\cal O}^{mat}_{\tree}$ is
given. Choosing instead of $x_N$ any other reference point $x_i \,
,$ we obtain a family ${\mathfrak{h}}^{mat}_{\tree}(x_i)$ of Lie
subalgebras. The restrictions of the above representation to these
subalgebras are all given by Theorem \ref{pierwsze}. In
particular, for $x_1 \, ,$ the corresponding Lie subalgebra
${\mathfrak{h}}^{mat}_{\tree}(x_1)$ is generated by the following
observables:
\begin{eqnarray}
 \widetilde Q_k & := & \hat {\mcalL}_{1k} \ , \\
 \widetilde P_k & := & \hat {\mcalM}_{1k} \ ,\\
 \widetilde R   & := & \hat {\mcalL}_{11} \ , \\
 \widetilde K & := & \mbox{\rm Re} \ (\hat {\mcalM}_{11}) \ ,
\end{eqnarray}
for $k = 2,\dots , N \, .$ The corresponding positive
representation is equivalent to the following one:

\begin{eqnarray}\label{trepr1}
  (\widetilde Q_k \widetilde \Psi)(\mu, w_2,\dots ,w_N)
  &=& e^\mu w_k
  \widetilde \Psi(\mu, w_2,\dots ,w_N) \ , \\
  \label{trepr2}
  (\widetilde P_k \widetilde \Psi)(\mu, w_2,\dots ,w_N)
  &=&  e^\mu
  \frac {2\hbar}{i} \frac {\partial}
  {\partial {\bar w}_k} \widetilde
  \Psi (\mu, w_2,\dots ,w_N)\ , \\
  \label{trepr4}
  (\widetilde R \widetilde \Psi)
  (\mu, w_2,\dots ,w_N) &=&
  e^{2\mu} \widetilde \Psi (\mu, w_2,\dots ,w_N) \ , \\
  \label{trepr3}
  (\widetilde K \widetilde \Psi)(\mu, w_2,\dots ,w_N)
  &=& \frac {\hbar}{i}
  \frac {\partial}
  {\partial \mu} \widetilde
  \Psi (\mu, w_2,\dots ,w_N) \ ,
\end{eqnarray}
with $\widetilde \Psi \in \widetilde{\cal H} \cong L^2 ({\mathbbm
R}^1 \times  {\mathbbm C}^{N - 1}) \, .$ We stress that if we
dealt with the restrictions of an arbitrary representation of the
Lie algebra ${\mathfrak{g}}^{mat}_{\tree}$ to the two subalgebras
under consideration, then these restrictions would be completely
independent. Here, the additional constraints (\ref{LLLL}) --
(\ref{RRRR}) imply a relation between the two Lie algebra
representations, which enables us to express the wave function
$\widetilde \Psi$ in terms of the wave function $\Psi$ {\em
uniquely}. For this purpose, take the polar decomposition of $w_N
\, ,$
\[
w_N = e^\lambda \xi \ ,
\]
where $|\xi |=1$ and $e^\lambda = |w_N| \, .$
For any $\Psi \in {\cal H} \, ,$ define the following
function $\Phi \in \widetilde{\cal H}$:
\begin{equation}\label{Phi}
  \Phi (\mu, w_2,\dots ,w_N) := {e^{\mu  - \lambda}} \, \Psi
  (e^\mu \xi^{-1} , w_2 \xi^{-1} , \dots , w_{N-1} \xi^{-1}, \lambda ) \
  .
\end{equation}
It is easy to check that this formula defines an isometric
isomorphism from ${\cal H}$ to $\widetilde{\cal H} \, .$ (The
factor in front of $\Psi$ is necessary to convert the radial
measure $|z_1|$d$|z_1|$ coming from the two dimensional measure
d$z_1$d${\bar z}_1$ into the measure d$\mu$ and the measure
d$\lambda$ into the radial measure $|w_N|$d$|w_N|$, coming from
the two dimensional measure d$w_N$d${\bar w_N}$ .) Hence, the
transformation from $\Phi$ to $\widetilde\Psi$ must be unitary,
too. We are going to prove that this transformation is of a
special type, consisting in multiplication by a phase factor only.

\begin{lemma}
\label{equi} Given a faithful, irreducible, positive and
non-degenerate representation of ${\cal O}^{mat}_{\tree} \, ,$ the
wave functions $\widetilde \Psi$ and $\Phi$ differ by a phase
factor depending only upon the phase $\xi$ of $w_N$. More
precisely, for any quantum state $\Psi$ we have:
\begin{equation}
\label{PhiPsi}
  \widetilde \Psi = \xi^{\frac Qe} \Phi \, ,
\end{equation}
where $Q$ is the total charge carried by the representation under
consideration.
\end{lemma}

\noindent {\bf Proof:} For $k=2,\dots , N-1 \, ,$ we have in the first
representation:
\begin{equation}\label{position1}
  R {\bar z_1} z_k = Q^*_1 Q_k = \hat {\mcalL}_{1N} \hat {\mcalL}_{Nk}
  = \hat {\mcalL}_{NN} \hat {\mcalL}_{1k} = R \widetilde
  Q_k\ ,
\end{equation}
which, due to positivity of $R$ implies $\widetilde Q_k = {\bar
z_1} z_k$. Thus, $\widetilde Q_k$ acts on $\Phi$ in the same way as on
$\widetilde \Psi:$
\[
\widetilde  Q_k = {\bar z_1} z_k = e^\mu \cdot \xi \cdot w_k \cdot
\xi^{-1} = e^\mu w_k \ .
\]
For $k = N \, ,$ this also holds true:
\[
\widetilde  Q_N = \hat {\mcalL}_{1N} = \left( \hat {\mcalL}_{N1}
\right)^* = Q^*_1 = {\bar z_1} e^\lambda = e^\mu \xi e^\lambda =
e^\mu w_N \ .
\]
Moreover, in the first representation we have :
\begin{equation}
  R \widetilde R = R \hat {\mcalL}_{11} = \hat {\mcalL}_{NN}
    \hat {\mcalL}_{11} = \left( \hat {\mcalL}_{N1} \right)^*
    \hat {\mcalL}_{N1}
     = Q^*_1 Q_1 = e^{2\lambda} {\bar z_1} z_1
    = R e^{2\mu} \ , \nonumber
\end{equation}
and, consequently, $\widetilde R = e^{2\mu}$. We conclude that the
``position operators'' $({\widetilde R}, {\widetilde  Q_k})$ of
the second representation act in the same way on $\Phi$ as on
$\widetilde \Psi$. This implies that $\Phi$ and $\widetilde \Psi$
may differ only by a phase factor,
\begin{equation}\label{faktor}
  \widetilde \Psi = \exp (i f) \Phi \ ,
\end{equation}
where $f$ is real. Next, we prove that the phase factor $f$ does
not depend upon variables $\mu \, ,$  $|w_N|$ and $w_k \,,$ for
$k=2,\dots ,N-1 \, .$ For this purpose, observe that in the first
representation we have
\[
 R \hat {\mcalM}_{11} = \hat {\mcalL}_{NN} \hat {\mcalM}_{11} =
 \left( \hat {\mcalL}_{N1} \right)^* \hat {\mcalM}_{N1} = Q^*_1 P_1
 = R {\bar z_1}   \cdot \frac {2\hbar}{i} \frac {\partial}
  {\partial {\bar z_1}} \ ,
\]
and, consequently
\begin{equation}\label{M-rozniczk}
  \hat {\mcalM}_{11} =  {\bar z_1}   \cdot
  \frac {2\hbar}{i} \frac {\partial}
  {\partial {\bar z_1}} \ .
\end{equation}
The real part of this operator applied to $\Phi$ gives
\begin{equation}\label{K-rozn}
  \widetilde K =\frac {\hbar}{i} \frac {\partial}
  {\partial \mu} \ ,
\end{equation}
which coincides with $\widetilde K$ acting on $\widetilde \Psi$.
We conclude that the phase factor cannot depend upon $\mu$. To
prove that it does not depend upon $w_k$, for $k = 2,\dots , N-1$,
neither, observe that in the first representation we have
\[
 R \widetilde P_k = \hat {\mcalL}_{NN} \hat {\mcalM}_{1k} = \left(
 \hat {\mcalL}_{N1} \right)^* \hat {\mcalM}_{Nk} = Q^*_1 P_k = R
 {\bar z_1}    \cdot \frac {2 \hbar}{i} \frac {\partial}
  {\partial {\bar z_k}} \ ,
\]
and, consequently,
\begin{equation}\label{tildePk1}
  \widetilde P_k =  {\bar z_1}   \cdot
  \frac {2 \hbar}{i} \frac {\partial}
  {\partial {\bar z_k}} \ .
\end{equation}
When applied to $\Phi$ it yields
\begin{equation}
\label{tildePk2}
  \widetilde P_k =  e^\mu \xi     \cdot
  \frac {2 \hbar}{i} \frac {\partial}
  {\partial {\bar z_k}} =
  e^\mu \cdot
  \frac {2 \hbar}{i} \frac {\partial}
  {\partial {\bar w_k}}\ ,
\end{equation}
exactly as in the $\widetilde \Psi$-representation. This implies
that $f$ does not depend upon $w_k$. Finally, to prove the
independence of $f$ from $|w_N|$ observe that in the second
representation holds
\begin{equation}
  \widetilde R \hat {\mcalM}_{NN} =  \hat {\mcalL}_{11} \hat
  {\mcalM}_{NN} = \left( \hat {\mcalL}_{1N} \right)^* \hat
  {\mcalM}_{1N}
  =  \widetilde Q^*_N \widetilde P_N = \widetilde R
  {\bar w_N}\cdot
  \frac {2 \hbar}{i} \frac {\partial}
  {\partial {\bar w_k}} \ ,\nonumber
\end{equation}
and, consequently,
\begin{equation}\label{Mfinal}
  \hat {\mcalM}_{NN} = {\bar w_N}\cdot
  \frac {2 \hbar}{i} \frac {\partial}
  {\partial {\bar w_N}} \ .
\end{equation}
The real part of this operator applied to $\widetilde \Psi$ yields
\begin{eqnarray}
  K & = & \mbox{\rm Re} \ (\hat {\mcalM}_{NN}) =
  \mbox{\rm Re} \left( {\bar w_N}\cdot
  \frac {2 \hbar}{i} \frac {\partial}
  {\partial {\bar w_N}} \right)  = \frac {\hbar}{i} \frac {\partial}
  {\partial \ |w_N|} |w_N|
  \nonumber \\
  & = & \frac {1}{|w_N|} \frac {\hbar}{i} \frac {\partial}
  {\partial \ \ln |w_N|} |w_N| \ . \label{Kfinal}
\end{eqnarray}
On the other hand, in the first representation $K$ acts on $\Psi$
as $\frac {\hbar}{i} \frac {\partial} {\partial \lambda}$.
Transformed to the representation in terms of $\Phi$ it gives
precisely (\ref{Kfinal}). This implies that the phase factor $f$
may depend only upon the phase of $w_N$, $f = f(\xi )$. To prove
the specific form of $f$ in equation (\ref{PhiPsi}), we
differentiate $\Phi$ in formula (\ref{Phi}) and get
\begin{equation}
  \frac ei \frac {\partial}{\partial \  \mbox{\rm arg} w_N} \Phi
  =
  \frac ei \frac {\partial}{\partial \  \mbox{\rm arg} \xi} \Phi
  =
  - \frac ei \sum_{k=1}^{N-1}
  \frac {\partial}{\partial \  \mbox{\rm arg} z_k} \Psi \ .
\end{equation}
Thus, by Lemma \ref{charge1} we have:
\begin{equation}\label{argwN}
  \frac ei \frac {\partial}{\partial \  \mbox{\rm arg} w_N} \Phi
  = -  \sum_{k=1}^{N-1} \hat q_{x_k} \Phi =
    (\hat q_{x_N} - Q)\Phi \ .
\end{equation}
On the other hand, in the second representation we have:
\begin{eqnarray}\label{chargeN2}
  \hat q_{x_N} \widetilde\Psi &=&
  \frac ei \frac {\partial}{\partial \  \mbox{\rm arg} w_N}
  \widetilde\Psi = \frac ei \frac {\partial}{\partial \  \mbox{\rm arg} \xi}
  \left( \exp (if) \Phi \right)
  \\
  &=& \exp (if) \left( e
  \frac {\partial f}{\partial \  \mbox{\rm arg} \xi} \Phi +
  (\hat q_{x_N} -  Q)\Phi
  \right) \ ,
\end{eqnarray}
which implies
\begin{equation}\label{phase-1}
  \frac {\partial f}{\partial \  \mbox{\rm arg} \xi} = \frac Qe \
  .
\end{equation}
\qed

As a consequence we get the following

\begin{theorem}
Given a value $Q \in e {\mathbbm N}$ of the total electric charge
carried by the matter field, there exists exactly one faithful,
irreducible, positive and non-degenerate representation of the
algebra ${\cal O}^{mat}_{\tree} \, ,$  assigned to a given tree.
This representation is uniquely generated by its restriction to
the Lie subalgebra ${\mathfrak{h}}^{mat}_{\tree}$. For different
values of $Q$ these representations are inequivalent. There is no
non-degenerate representation of ${\cal O}^{mat}_{\tree}$
corresponding to $Q \notin e {\mathbbm N}$.
\end{theorem}

{\bf Proof:} Choosing different points $x_k$, together with the
corresponding Lie subalgebras ${\mathfrak{h}}^{mat}_{\tree}(x_k)$,
and transforming between different representations, we may
uniquely calculate elements $\hat {\mcalL}_{kl}$, $\hat
{\mcalM}_{kl}$ and $\hat {\mcalN}_{kl} = \left( \hat
{\mcalM}_{lk}\right)^*$, for $k \ne l$, in a single, fixed
representation, say the first one, described by the wave function
$\Psi$. Then, we can represent also all the ``ultralocal''
observables $\hat {\mcalL}_{kk}$, $\hat {\mcalM}_{kk}$ and $\hat
{\mcalN}_{kk} = \left( \hat {\mcalM}_{kk}\right)^*$ in this fixed
representation. Indeed, in the representation based on $x_k$ we
have $\hat {\mcalL}_{kk}=R$, whereas
\begin{equation}\label{Mxx}
  \hat {\mcalM}_{kk} = K + i \hbar \left(
  \frac {\hat q_{x_k}}e + {\mathbbm 1} \right)
\end{equation}
according to formulae (\ref{ReM}) and (\ref{ladunek1}). Finally,
we find the representation of the observables $\hat {\mcalR}_{kl}$
by choosing any $x_n \, ,$ different from both $x_k$ and $x_l \,
,$ and using formula
\begin{equation}\label{Rkl}
  \hat {\mcalL}_{nn}\hat {\mcalR}_{kl}=
  \hat {\mcalN}_{kn} \hat {\mcalM}_{nl} \ .
\end{equation}
\qed

We denote the irreducible representation space corresponding to
eigenvalue $Z$ of the total charge $\tfrac{1}{e} \hat Q $ by
${\cal H}^{mat}_{\tree}(Z)\, .$

\subsection{Charge superselection structure}

Now we use the fact that, for a given tree, the observable algebra
decomposes,
\begin{equation}
{\cal O}(\Lambda ) = {\cal O}^{em}_{\tree} \otimes
{\cal O}^{mat}_{\tree} \ ,
\end{equation}
see (\ref{tensor}). But the electromagnetic part ${\cal
O}^{em}_{\tree}$ is generated by a finite number of canonically
conjugate pairs. Thus, its integrable representations are (up to
unitary equivalence) unique, due to the von Neumann theorem
\cite{vN}. Thus, the faithful, irreducible and non-degenerate
representations of ${\cal O}(\Lambda )$ are labeled by the
irreducible charge sectors of the matter field part ${\cal
O}^{mat}_{\tree}$ , as described in the previous subsection. For a
given tree, we get the physical Hilbert space as a direct sum of
charge superselection sectors,
\begin{equation}
\label{superselection}
   {\cal H}(\Lambda ) = \bigoplus _Z \left\{{\cal H}^{em}_{\tree} \otimes
   {\cal H}^{mat}_{\tree}(Z)\right\} \ .
\end{equation}

Finally, it can be easily shown that a different choice of the
tree induces a similar decomposition of ${\cal O}(\Lambda ) $ as
above and the two decompositions are related to each other via an
isomorphism of the corresponding electromagnetic observable
algebras.


\section{Discussion}


As we have seen, the fact that the observable algebra is generated
by a certain Lie algebra, is extremely helpful for the
classification of its irreducible representations. It should be
also helpful for constructing the thermodynamic limit, because for
taking the limit $N \rightarrow \infty$ in the generating Lie
algebra there seems to exist appropriate mathematical tools for
studying the resulting representations, see \cite{Kac}. However,
it is doubtful, whether the classification of charge
superselection sectors obtained here will carry over to the
thermodynamic limit in a straightforward way. One should rather
expect that these considerations will have to be supplemented by a
discussion of field dynamics. For this purpose we define the
lattice version of the field Hamiltonian (\ref{Ham}) by the
following procedure. In formulae (\ref{l-def}) -- (\ref{r-def}) we
put $\gamma = (x,x+\hat k)$ and use the substitution
\begin{eqnarray}\label{approxA}
  \exp(i g \int_{x}^{x+\hat k} \hat A) & \simeq &
  {\hat {\mathbbm{1}}} + i a g A_k \ ,
  \\
  {\hat \phi}_{x+\hat k} & \simeq &
  {\hat \phi}_{x} + a \partial_k {\hat \phi}_{x} \ .
  \label{approxB}
\end{eqnarray}
Consequently, we obtain the following approximation:
\begin{equation}\label{H-int}
  |\vec{D} \phi (x)|^2 \simeq \frac 1{a^2} (\hat{\mcalL}_{xx})^{-1}
  \sum_{\hat k} \left( \hat{\mcalL}_{(x,x+\hat k)} - \hat{\mcalL}_{xx}
  \right)^* \left( \hat{\mcalL}_{(x,x+\hat k)} - \hat{\mcalL}_{xx}
  \right) \ .
\end{equation}
This leads to the lattice version of the Hamiltonian
(\ref{Ham}):
\begin{equation}
 {\hat H} =  {\hat H}_{el-mag} + {\hat H}_{matter} + {\hat H}_{int}\ ,
 \label{H-quantum}
\end{equation}
where the electromagnetic, matter and interaction parts of the
Hamiltonian are given by:
\begin{eqnarray}
 {\hat H}_{el-mag} & = & \frac a2  \sum_{(x,x+\hat k)}
 \left( \hat E_{x,x+\hat k} \right)^2 +
 \frac a2 \sum_{x;\hat k,\hat l}
 \left( \hat B_{x;\hat k,\hat l} \right)^2 \ ,
 \\
 {\hat H}_{matter} & = & a^3 \sum_x \left( \frac 12
 \hat{\mcalR}_{xx} + V(\hat{\mcalL}_{xx}) \right) \ ,
\\
 {\hat H}_{int} & = &  \frac a2
 \sum_{x,\hat k} (\hat{\mcalL}_{xx})^{-1}
 \left( \hat{\mcalL}_{(x,x+\hat k)} - \hat{\mcalL}_{xx}
  \right)^* \left( \hat{\mcalL}_{(x,x+\hat k)} - \hat{\mcalL}_{xx}
  \right) \ .
\end{eqnarray}
Given a finite lattice, we have thus approximated the field by a
quantum-mechanical system whose dynamics is governed by the
positively defined Hamiltonian\footnote{When considering different
boundary conditions, an additional surface term in the Hamiltonian
will be necessary.} (\ref{H-quantum}). Its ground state will be
treated as a finite-lattice-approximation of the vacuum. Suppose
that this vacuum state has been found. Then, applying the strategy
outlined in \cite{qedspin}, one should find the vacuum in the
thermodynamic limit as a projective limit of the above vacuum
states corresponding to finite lattices. Using this vacuum, the
physically admissible representations of the observable algebra
(in the thermodynamic limit) could be singled out via the
GNS-construction. To find the ``true'' vacuum will be, of course,
extremely difficult, but may be one can find an approximation,
which is ``much better'' than the perturbative vacuum. Then one
can hope to get deeper insight into nonperturbative aspects of
this model. The same remark applies to one- or multi-particle
states.

Of course, the ultra-violet limit of the theory (obtained by
sending the lattice spacing $a$ to zero) is, probably, much more
difficult to investigate. But, in principle, the strategy outlined
in \cite{qedspin} applies also here.

\section*{Acknowledgments}

The authors are very much indebted to L.~S.~Woronowicz for helpful
discussions and remarks. One of the authors (J.~K.) is grateful to
Professor E.~Zeidler for his hospitality at the Max Planck
Institute for Mathematics in the Sciences, Leipzig, Germany.


\begin{appendix}

\section*{Appendix}
\renewcommand{\theequation}{\Alph{section}.\arabic{equation}}

\section{A Non-integrable Representation Carrying Non-inte\-ger Charge}
\label{Example}
\setcounter{equation}{0}

Quantization of charge is due to integrability of the
representations of the Lie algebra generating the matter field
part of the observable algebra. Below, we construct a weak
(non-integrable) representation, which carries non-integer charge.
At the same time we construct an interesting example of weakly
commuting operators, which do not commute strongly (cf.
\-\cite{nels}).

We limit ourselves to a single lattice point ($N = 1$) and
multiply the wave function by the phase factor $\xi^{c}= \exp (i c
\cdot \mbox{\rm arg} z) $, $c \in {\mathbbm N} \, ,$ as in formula
(\ref{PhiPsi}). After this operation the spectrum of the  charge
operator
\begin{equation}\label{ang-mom}
 \hat j := \frac {\hat q}e =
  \frac 1i \frac {\partial}{\partial \  \mbox{\rm arg} z} \ ,
\end{equation}
gets shifted by the value $c$. It is interesting that a similar
``shift'' can be also defined for any real number $c \in {\mathbbm
R}$. Of course, the result of such a ``shift'' cannot be unitarily
equivalent to the original operator $\hat j$. As will be seen in
the sequel, this leads to a simple example of canonical
commutation relations, which are fulfilled only in the weak sense:
momenta $\hat p_1$ and $\hat p_2$ do not commute strongly or, in
other words, $\hat P =\hat p_1 +i \hat p_2$ is not normal.

Consider, therefore, the Hilbert space ${\cal H} = L^2 ({\mathbbm C})$
and define the following two groups of unitary transformations:
\begin{eqnarray}\label{U12}
  (\hat U_1(t) \Psi )(z) & = & \frac 1{\varphi_1 (z)}
  (\varphi_1 (z + t \hbar ) \Psi (z + t \hbar )) \ , \\
  (\hat U_2(s) \Psi )(z) & = & \frac 1{\varphi_2 (z)}
  (\varphi_2 (z + i s \hbar ) \Psi (z + i s \hbar )) \ ,
  \label{U2}
\end{eqnarray}
where $\varphi_1$ (respectively $\varphi_2$) is a phase factor
obtained from the multivalued function
\[
\varphi (z) := \exp (i c \cdot \mbox{\rm arg} z) \ ,
\]
by the cut along the real (respectively the imaginary) positive
half-axis. Denote by $\hat p_a$, $a=1,2$, the self-adjoint
generators of these groups:
\[
\hat U_a(t) = \exp ( i t p_a ) \ .
\]
These operators may be easily calculated on their common, dense
domain $D$, consisting of those functions, which are smooth and
vanish identically in a neighbourhood of the two cuts. Indeed, for
$\Psi \in D$, a straightforward calculation gives:
\begin{eqnarray}\label{P12}
  (\hat p_1 \Psi )(z) & = & \frac 1{\varphi_1 (z)} \frac {\hbar}i
  \frac {\partial}  {\partial \  {\mbox{\rm Re}\ z}}
  (\varphi_1 (z) \Psi (z)) \ , \\
  (\hat p_2 \Psi )(z) & = & \frac 1{\varphi_2 (z)} \frac {\hbar}i
  \frac {\partial}  {\partial \  {\mbox{\rm Im}\ z}}
  (\varphi_2 (z) \Psi (z)) \ .
\end{eqnarray}
This immediately implies weak commutation:
\[
\hat p_1 \hat p_2 \Psi =  \hat p_2 \hat p_1 \Psi \ .
\]
However, the commutation relations are not satisfied in the strong
sense, because the groups $\hat U_1$ and $\hat U_2$ do not
commute. Indeed, take any smooth function $\Psi$ with support
contained in the square $S = \{ -2 \le {\mbox{\rm Re}\ z},
{\mbox{\rm Im}\ z} \le -1 \}$. It is easy to check that we have
\begin{equation}\label{non-commm}
   \hat U_2(3) \hat U_1(3) \Psi = \exp (2\pi i c)
   \hat U_1(3) \hat U_2(3) \Psi \ ,
\end{equation}
and, whence, we obtain the strong commutation only for $c \in
{\mathbbm N}$. A simple geometric interpretation of this result
consists in interpreting the multivalued function $\Phi=\varphi
\cdot \Psi$ as a function defined on the Riemann surface ${\cal
R}$ of the logarithm,
\[
 {\cal R} = \{ ( |z| , \mbox{\rm arg}z ): |z| \in {\mathbbm R}_+ \,
 ,\,
 \mbox{\rm arg}z \in {\mathbbm R}^1  \} \ ,
\]
and satisfying the condition:
\[
\Phi (|z| ,\mbox{\rm arg} z + 2\pi ) = \exp (2\pi i c) \Phi (|z|
,\mbox{\rm arg} z) \ .
\]
These functions form a Hilbert space $\widetilde{\cal H}$ with
scalar product defined by integration over any closed set $D
\subset {\cal R}$, which covers  almost the whole space ${\mathbbm
C}$ only once. The mapping ${\cal H} \ni \Psi \rightarrow \Phi :=
\varphi \Psi \in \widetilde{\cal H}$ is an isomorphism of Hilbert
spaces. When interpreted in terms of $\widetilde{\cal H}$, the
operators $\hat U_2(3) \hat U_1(3)$ and $\hat U_1(3) \hat U_2(3)$
describe dragging $\Phi$ along the spiral surface ${\cal R}$ with
respect to the two opposite helicities.

Now, consider the operator $\hat X = \hat x_1 + i \hat x_2$, where
$\hat x_1$ (resp. $\hat x_2$) is the operator of multiplication by
the real (resp. the imaginary) part of $z$. Formulae (\ref{U12})
and (\ref{U2}) imply that $\hat p_1$ and $\hat x_1$ satisfy the
canonical commutation relations in the strong sense. The same is
true for $\hat p_2$ and the operator $\hat x_2$. Hence, formally
$\hat P =\hat p_1 + i \hat p_2$ and $\hat X = \hat x_1 + i \hat
x_2$ satisfy the commutation relations (\ref{phip-lattice}) for
quantum mechanics of two degrees of freedom in the weak sense.
But, these relations are not satisfied strongly, because the real
and the imaginary parts of $\hat P$, although being self-adjoint,
do not commute strongly. It is easy to show that the formal
``charge operator'' built from them,
\[
\hat q := {\mbox{\rm Im}} \left( \hat X^* \hat P \right) \ ,
\]
is self-adjoint and has spectrum shifted by $c$ with respect
to the ordinary spectrum:
\[
\mbox{\rm Sp}\  \hat q = \{ n + c : n \in {\mathbbm N}  \} \ .
\]
We conclude that the operators $\hat{\cal L}=\hat X^* \hat X$,
$\hat{\cal M}=\hat X^* \hat P$, $\hat{\cal N}=\hat P^* \hat X$ and
$\hat{\cal R}=\hat P^* \hat P$ satisfy the axioms of the
observable algebra weakly, but do not provide its strong
(integrable) realization.

\section{A Unified Description of the Bosonic and the Fermionic
Case}
\label{Unification}
\setcounter{equation}{0}

Consider at each lattice point $x_k$ the following bosonic
annihilation and creation  operators:
\begin{eqnarray}\label{ak}
  a_k &:=& \tfrac 1 {\sqrt{2}} \left( {\rm Re} \hat \phi_k
  + \tfrac i\hbar {\rm Re} \hat \pi_k \right) \ ,
  \\
  b_k &:=& \tfrac 1 {\sqrt{2}} \left( {\rm Im} \hat \phi_k
  + \tfrac i\hbar {\rm Im} \hat \pi_k \right) \ ,
  \\
  a^*_k &:=& \tfrac 1 {\sqrt{2}} \left( {\rm Re} \hat \phi_k
  - \tfrac i\hbar {\rm Re} \hat \pi_k \right) \ ,
  \\
  b^*_k &:=& \tfrac 1 {\sqrt{2}} \left( {\rm Im} \hat \phi_k
  - \tfrac i\hbar {\rm Im} \hat \pi_k \right) \ .
\end{eqnarray}
Then, take their combinations:
\begin{eqnarray}\label{chi}
  \chi_k &:=& \tfrac 1 {\sqrt{2}}
  \left( a_k + i b_k \right)
  = \tfrac 12 \left( \hat \phi_k +
  \tfrac i\hbar  \hat \pi_k \right)\ ,
  \\
  \varphi^*_k &:=& \tfrac 1 {\sqrt{2}}
  \left( a^*_k + i b^*_k \right)
  = \tfrac 12 \left( \hat \phi_k -
  \tfrac i\hbar  \hat \pi_k \right)\ ,
  \\
  \chi^*_k &:=& \tfrac 1 {\sqrt{2}}
  \left( a^*_k - i b^*_k \right)
  = \tfrac 12 \left( \hat \phi^*_k -
  \tfrac i\hbar  \hat \pi^*_k \right)\ ,
  \\
  \varphi_k &:=& \tfrac 1 {\sqrt{2}}
  \left( a_k - i b_k \right)
  = \tfrac 12 \left( \hat \phi^*_k +
  \tfrac i\hbar  \hat \pi^*_k \right) \ .
\end{eqnarray}
The entire information about the field algebra may be encoded in
the following objects:
\begin{equation}\label{psi}
 \psi_k = \left( \begin{array}{c} \chi_k \\
 \varphi^*_k \end{array} \right)  \ ,
 \ \ \ \ \ \ \ \
 \psi^{*}_k = \left( \begin{array}{c} \chi^{*}_k \\
 \varphi_k \end{array} \right) \ .
\end{equation}
In \cite{qedspin} and \cite{qedspin1} we considered spinoral QED,
where the matter field was described by a similar structure. The
only difference was, that there both $\chi$ and $\varphi$ carried
an additional spinorial index $K = 1,2$, (but this only multiplies
the number of degrees of freedom). These objects fulfill the
canonical (anti)-commutation  relations:
\begin{equation}\label{canon+}
  [\chi_k , \chi^*_l ]_{\mp} =
  [\varphi_k , \varphi^*_l ]_{\mp} = \delta_{kl}
  \hat{\mathbbm 1}    \ ,
\end{equation}
where the upper sign always applies to the bosonic and the lower
to the fermionic case. The remaining (anti)-commutators vanish.
Due to (\ref{gauge1}), under a gauge transformation the field
$\psi$ is multiplied by $ e^{-i g \lambda(x)}$, whereas $\psi^*$
is multiplied by $ e^{i g \lambda(x)}$. Hence, as the observable
algebra generators we may use the following gauge invariant
combinations:
\begin{eqnarray}\label{lij}
  l_{ij} &:=& \pm \varphi_i
  \exp(i g \int_\gamma \hat A) \
  \varphi_j^* \ ,
  \\
  r_{ij} &:=& - \chi^*_i
  \exp(i g \int_\gamma \hat A) \
  \chi_j \ ,
  \\
  m_{ij} &:=& i^{\epsilon }\varphi_i
  \exp(i g \int_\gamma \hat A) \
  \chi_j \ ,
  \\
  n_{ij} &:=& i^{\epsilon }\chi_i^*
  \exp(i g \int_\gamma \hat A) \
  \varphi^*_j  \ .
\end{eqnarray}
Here, by $\epsilon$ we denote the number $\epsilon : = \frac{1 \mp
1}{2}$, which equals $0$ for the bosonic and $1$ for the fermionic
case. It is easy to check that these generators fulfill the
following {\em universal} commutation relations, the same for
bosons as for fermions:
\begin{eqnarray}\label{ll}
  \left[ l_{ij} , l_{kl} \right] & = &
  - \delta_{kj} l_{il} + \delta_{il} l_{kj} \, ,\\
  \left[ l_{ij},m_{kl} \right] & = &
   - \delta_{kj} m_{il}   \ ,
  \\
  \left[ l_{ij}, n_{kl} \right] & = &
   \delta_{il} n_{kj}   \ ,
  \\
  \left[ l_{ij}, r_{kl} \right] & = &
  0   \ ,
  \\
  \left[ m_{ij} , m_{kl} \right] & = &
  0   \ ,
  \\
  \left[ m_{ij} , n_{kl} \right] & = &
  \delta_{kj} l_{il} - \delta_{il} r_{kj}   \ ,
  \\
  \left[ m_{ij}, r_{kl} \right] & = &
   - \delta_{kj} m_{il}   \ ,
   \\
  \left[ n_{ij} , n_{kl} \right] & = &
  0   \ ,
  \\
  \left[ n_{ij}, r_{kl} \right] & = &
  \delta_{il} n_{kj}  \ ,
  \\
  \left[ r_{ij} , r_{kl} \right] & = &
  - \delta_{kj} r_{il} + \delta_{il} r_{kj}   \ .
\end{eqnarray}
But the conjugation is different in both cases:
\begin{equation}\label{star-boz}
  l^*_{ij} = l_{ji} \ , \ \ \ r^*_{ij} = r_{ji} \ , \ \ \
  m^*_{ij} = \pm n_{ji} \ .
\end{equation}
It is easy to check that, under the matrix presentation
(\ref{identyfikacja1}) and (\ref{identyfikacja2}), we have:
\begin{eqnarray}
 l_{kl} & := & \left(
  \begin{smallmatrix} iE_{kl}&
  0\\ 0& 0\end{smallmatrix} \right)
  \ \ \ \ \ \ \,
 m_{kl} := i^\epsilon
  \left( \begin{smallmatrix} 0&
  E_{kl}\\ 0&0\end{smallmatrix} \right)
  \label{identyfikacja11} \\
 n_{kl} & := & i^\epsilon \left(
  \begin{smallmatrix} 0&
  0\\ E_{kl}& 0\end{smallmatrix} \right)
  \ \ \ \ \ \
 r_{kl} :=
  \left( \begin{smallmatrix} 0&
    0\\ 0& iE_{kl}\end{smallmatrix} \right)
  \ .\label{identyfikacja12}
\end{eqnarray}
In the bosonic case, the conjugation (\ref{star-boz}) implies
(\ref{conjugationgl}) and, therefore, the algebra of self-adjoint
observables is generated by $u(N,N)$.  In the fermionic case,
(\ref{star-boz}) implies $A^* = - A^\dag$ and the algebra of
self-adjoint observables is generated by two copies of $u(2N)$,
corresponding to two values of the spinorial index $K$. In both
cases, the following formula for the total charge holds:
\begin{equation}
 \tfrac 1e \  Q = \sum l_{ii} + \sum r_{ii} - \mathbbm{1} \ .
\end{equation}

\end{appendix}

\end{document}